\documentclass[epj,nopacs]{svjour}
\pdfoutput=1

\usepackage{preprintcover}  
\usepackage{authblk}
\usepackage{atlasphysics}
\usepackage{rotating}
\usepackage{mathrsfs}
\usepackage{amsmath}
\usepackage{amssymb}
\usepackage{booktabs}
\usepackage{xspace}
\usepackage[colorlinks,breaklinks,pdftitle={Measurement of kT splitting scales in W->lv events at sqrt(s)=7 TeV with the ATLAS detector},pdfauthor={The ATLAS Collaboration},linkcolor=blue,citecolor=blue,filecolor=black,urlcolor=blue]{hyperref}
\usepackage{verbatim}
\usepackage{cite}
\newcommand\kt{\ensuremath{k_{\rm T}}\xspace}

\newcommand{\alpgen}{\textsc{Alpgen+Herwig}\xspace}
\newcommand{\sherpa}{\textsc{Sherpa}\xspace}

\newcommand{\mcatnlo}{\textsc{Mc\scalebox{0.8}{@}Nlo}\xspace}
\newcommand{\powhegsix}{\textsc{Powheg+Pythia6}\xspace}
\newcommand{\powhegeight}{\textsc{Powheg+Pythia8}\xspace}
\DeclareGraphicsExtensions{.pdf}

\PreprintCoverPaperTitle{Measurement of \kt splitting scales in $W\to \ell \nu$\\
  events at $\sqrt{s}$ = 7 TeV with the ATLAS detector}  
\PreprintIdNumber{CERN-PH-EP-2013-003}  
\PreprintCoverAbstract{A measurement of splitting scales, as defined by the \kt clustering algorithm, is presented for final states containing a $W$ boson produced in proton--proton collisions at a centre-of-mass energy of 7~TeV. The measurement is based on the full 2010 data sample corresponding to an integrated luminosity of 36~pb$^{-1}$ which was collected using the ATLAS detector at the CERN Large Hadron Collider. Cluster splitting scales are measured in events containing $W$ bosons decaying to electrons or muons. The measurement comprises the four hardest splitting scales in a \kt cluster sequence of the hadronic activity accompanying the $W$ boson, and ratios of these splitting scales. Backgrounds such as multi-jet and top-quark-pair production are subtracted and the results are corrected for detector effects. Predictions from various Monte Carlo event generators at particle level are compared to the data. Overall, reasonable agreement is found with all generators, but larger deviations between the predictions and the data are evident in the soft regions of the splitting scales.
}  
\PreprintJournalName{EPJC}  
\begin{document}
\title{Measurement of \kt splitting scales in $W\to \ell \nu$\\
  events at $\sqrt{s}$ = 7 TeV with the ATLAS detector}
\author{The ATLAS Collaboration}
\titlerunning{Measurement of \kt splitting scales in $W\to \ell \nu$ events at $\sqrt{s}$ = 7 TeV}
\authorrunning{ATLAS}
\institute{}
\date{Received: date / Revised version: date}
%
\abstract{A measurement of splitting scales, as defined by the \kt clustering algorithm, is presented for final states containing a $W$ boson produced in proton--proton collisions at a centre-of-mass energy of 7~TeV. The measurement is based on the full 2010 data sample corresponding to an integrated luminosity of 36~pb$^{-1}$ which was collected using the ATLAS detector at the CERN Large Hadron Collider. Cluster splitting scales are measured in events containing $W$ bosons decaying to electrons or muons. The measurement comprises the four hardest splitting scales in a \kt cluster sequence of the hadronic activity accompanying the $W$ boson, and ratios of these splitting scales. Backgrounds such as multi-jet and top-quark-pair production are subtracted and the results are corrected for detector effects. Predictions from various Monte Carlo event generators at particle level are compared to the data. Overall, reasonable agreement is found with all generators, but larger deviations between the predictions and the data are evident in the soft regions of the splitting scales.
}
\maketitle
\section{Introduction}
\label{sec:intro}
The CERN Large Hadron Collider (LHC), in addition to being a discovery machine, produces
a wealth of data suitable for studies of the strong interaction.
Due to the strongly interacting partons in the initial state and the
large phase space available, final states often include hard jets arising from QCD brems\-strahlung. Discovery signals, on the other hand, often
contain jets from quarks produced in electroweak interactions.
A robust understanding of QCD-initiated processes in measurement and theory is necessary in order to distinguish such signals from backgrounds.

One critical background for searches is the $W$+jets process in the leptonic decay mode, which
provides a large amount of missing transverse momentum together with
jets and a lepton. This process is a testing ground for recent
progress in QCD calculations, e.g. at fixed order~\cite{Berger:2010zx,Ellis:2008qc}
or in combination with resummation~\cite{Frederix:2011ig,Hoche:2010kg,Hoeche:2012ft},
and it has been measured using many observables at both the
Tevatron~\cite{Aaltonen:2007ip,Abazov:2011rf}
and the LHC~\cite{Chatrchyan:2011ig,Chatrchyan:2011ne,Chatrchyan:2012he,Aad:2011xn,Aad:2011kp,Aad:2012en,ATLAS:2012au}.

In this paper the \kt jet finding algorithm~\cite{Catani:1993hr,Ellis:1993tq} is employed for a measurement of
differential distributions of the \kt splitting scales in $W$+jets events.
These measurements aim to provide results which can be
interpreted particularly well in a theoretical context and
improve the theoretical modelling of QCD effects.
The measurement was performed independently in the electron ($\Wen$) and
muon ($\Wmn$) final states.
Backgrounds such as multi-jet and top-quark pair production were subtracted
and results were corrected for detector effects. 
The resulting data distributions are compared to predictions from various Monte Carlo event generators
at particle level.

After an outline of the measurement in this section, the data analysis and event selection are summarised in Sect.~\ref{sec:data}. The Monte Carlo (MC) simulations used for theory comparisons are described in Sect.~\ref{sec:mc}. Distributions at the detector level are displayed in Sect.~\ref{sec:DetLevelDists}. The procedure used to correct these to the particle level before any detector effects is outlined in Sect.~\ref{sec:particlelevel} together with a weighting technique used to maximise the statistical power available, whilst minimising the systematic uncertainty arising from pileup. The evaluation of the systematic uncertainties is summarised in Sect.~\ref{sec:unc}, and the results are shown in Sect.~\ref{section_results}, followed by the conclusions in Sect.~\ref{sec:conclusions}.

\subsection{Definition of \kt splitting scales}

\label{sec:variables}
\newcommand{\di}{\ensuremath{d_{k}}\xspace}

The \kt jet algorithm is a sequential recombination algorithm. Its
splitting scales are determined
by clustering objects together according to their distance from each other.
The inclusive \kt algorithm uses the following distance
definition~{\cite{Catani:1993hr,Ellis:1993tq}}:
\begin{align}
  & d_{ij} = \min(p_{\mathrm{T}i}^2,p_{\mathrm{T}j}^2)\frac{\Delta R_{ij}^2}{R^2}, \;
  \Delta R_{ij}^2=(y_i-y_j)^2+(\phi_i-\phi_j)^2, \nonumber \\
  & d_{iB} = p_{\mathrm{T}i}^2, \label{eq:dij}
\end{align}
where the transverse momentum $p_\mathrm{T}$, rapidity $y$ and azimuthal angle $\phi$ of
the input objects are labelled with an index corresponding to the
$i$-th and $j$-th momentum in the input configuration, and $B$ denotes a beam.
These momenta can be determined using energy deposits in the calorimeter at the
detector level, or hadrons at the particle level in Monte Carlo simulation.
The $R$ parameter was chosen to be $R=0.6$ in this paper, which is an intermediate choice between small values $R\approx 0.2$, whose narrow width minimizes the impact of pileup and the underlying event, and $R\approx 1.0$, whose large width efficiently collects radiation.

The clustering from the set of input momenta proceeds along the following
lines:
\begin{enumerate}
\item Calculate $d_{ij}$ and $d_{iB}$ for all $i$ and $j$ from the input momenta according to Eq.~\eqref{eq:dij}. \label{kt:step1}
\item Find their minimum. \label{kt:step2}
  \begin{enumerate}
  \item If the minimum is a $d_{ij}$, combine $i$ and $j$ into a single momentum in the list of input momenta: $p_{ij}=p_i+p_j$
  \item If the minimum is a $d_{iB}$, remove $i$ from the input momenta and declare it to be a jet.
  \end{enumerate}
\item Return to step~\ref{kt:step1} or stop when no particle remains.
\end{enumerate}

The observables measured are defined as the smallest of the
square roots of the $d_{ij}$ and $d_{iB}$ variables
($\sqrt{d_{ij}}$, $\sqrt{d_{iB}}$)
found at each step in the clustering sequence.
To simplify the notation they are commonly referred to as the splitting
scales $\sqrt{\di}$, which stand for the minima that
occur when the input list proceeds from $k+1$ to $k$ momenta by
clustering and removing in each step. For example, $\sqrt{d_{0}}$ is found from the last step in the clustering sequence and reduces to the transverse
momentum of the highest-$p_{\rm{T}}$ jet.

\begin{figure}[htb]
\begin{center}
\includegraphics[width=0.2\textwidth]{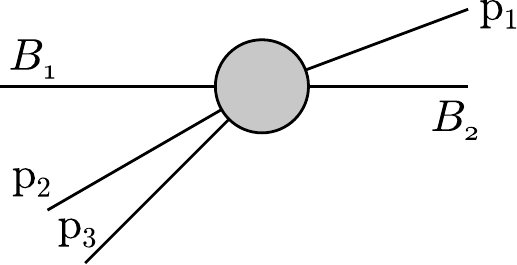}\\
\vspace*{-2mm}Step 0:\\Input momenta\vspace*{8mm}\\
\includegraphics[width=0.2\textwidth]{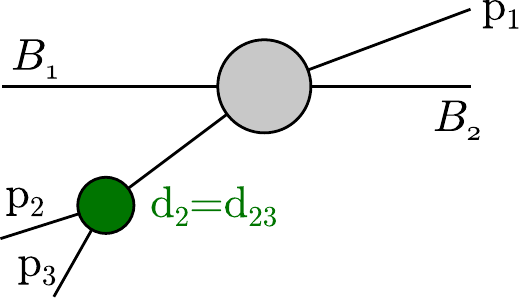}\\
\vspace*{-2mm}Step 1:\\Merge $3 \to 2$\vspace*{8mm}\\
\includegraphics[width=0.2\textwidth]{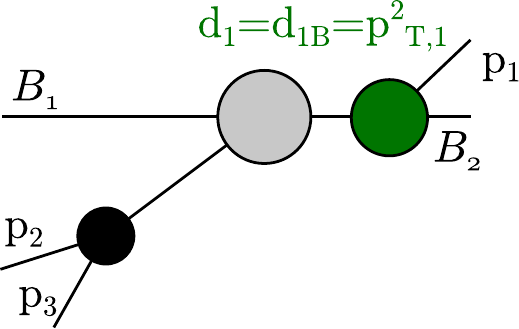}\\
\vspace*{-2mm}Step 2:\\Merge $2 \to 1$\vspace*{8mm}\\
\includegraphics[width=0.2\textwidth]{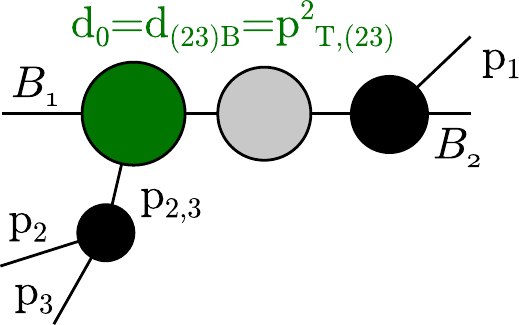}\\
\vspace*{-2mm}Step 3:\\Merge $1 \to 0$\vspace*{2mm}
\caption{Illustration of the \kt clustering sequence starting from the original input configuration (three objects $p_1$, $p_2$, $p_3$, and beams $B_1$, $B_2$). At each step, $k+1$ objects are merged to $k$.}
\label{fig:clustering}
\end{center}
\end{figure}

Figure~\ref{fig:clustering} schematically displays the clustering
sequence derived from an original input configuration of three objects labelled
$p_1$, $p_2$, $p_3$ in the presence of beams $B_1$ and $B_2$. In the first clustering
step, where three objects are grouped into two (denoted $3\to 2$),
the minimal splitting scale is found between momenta $p_2$ and $p_3$,
leading to $d_2=d_{23}$. In the second step ($2\to 1$), the momentum $p_1$ is closest
to the beam, and thus is removed and declared a jet at the scale
$d_1=d_{1B}=p_{\mathrm{T}1}^2$. Ultimately, the third clustering ($1\to 0$) has only
the beam distance of the combined input $p_{2,3}$ remaining, leading to a
scale of $d_0=d_{(23)B}=p_{\mathrm{T},(23)}^2$.

\subsection{Features of the observables}

An important feature of these observables is their separation
into two regions: a ``hard'' one with $\sqrt{\di}\gtrsim 20$~GeV which is
dominated by perturbative QCD effects, and a ``soft'' one in which more
phenomenological modelling aspects such as hadronisation and multiple
partonic interactions may exert substantial influence on theory predictions.
The number of events in the hard region for high $k$
is naturally low in the data sample analysed for this measurement.
Thus for statistical reasons values of $0\leq k\leq3$ are considered
in this publication.
No explicit jet requirement is imposed in the event selection.

In addition to the observables mentioned above,
it is also interesting to study ratios of consecutive clustering values, $\sqrt{d_{k+1}/d_k}$,
where some experimental uncertainty cancellations occur, as discussed in Sect.~\ref{sec:unc}.
Of particular interest is the region whe\-re $\sqrt{d_{k+1}/d_k}\to 1$, as it probes events with
subsequent emissions at similar scales. Those events could be challenging to
describe correctly for parton shower generators without matrix element corrections.
The splitting scale ratio amounts to a normalisation of the splitting scale to the scale
of the QCD activity in the ``underlying process'', i.e.~after the clustering.
To reduce the influence of non-perturbative effects, each ratio observable
$\sqrt{d_{k+1}/d_{k}}$ is measured with events satisfying $\sqrt{d_{k}}>20$~GeV.

The central idea underlying this measurement is that the
measure of the \kt algorithm corresponds relatively well to the
singularity structure of QCD. To illustrate this,
the small-angle limit of the squared \kt measure is given in terms of the
angle $\theta_{ij}$ between two momenta $i$ and $j$, and the energy corresponding to the softer momentum, $E_{i}$, by~\cite{Catani:1993hr}:
\begin{align}
  \label{eq:ktlimit}
  p_{\mathrm{T}i}^2 \Delta R_{ij}^2 &\simeq E_i^2 \theta_{ij}^2 \\
  p_{\mathrm{T}i}^2 &\simeq E_i^2\theta_{iB}^2\;,
\end{align}
while the splitting probability for a final-state branching into partons
$i$ and $j$ evaluates to
\begin{equation}
  \label{eq:qcddiv}
  \frac{\mathrm{d}P_{ij\to i,j}}{\mathrm{d}E_i\mathrm{d}\theta_{ij}} \sim \frac{1}{\min(E_i,E_j)\theta_{ij}}
\end{equation}
in the collinear limit~\cite{Salam:2009jx}.

From a comparison of Eqs.~\eqref{eq:ktlimit} and \eqref{eq:qcddiv} it can be
seen that each step of the \kt algorithm identifies the parton pair
which would be the most likely to have been produced by QCD interactions.
In that sense, this clustering
sequence mimicks the reversal of the QCD evolution.

In contrast the anti-$k_t$~\cite{Cacciari:2008gp}
algorithm cannot be used in the
same way: its distance measure replaces all $p_\mathrm{T}^2$ by $p_\mathrm{T}^{-2}$.
So even though collinear branchings are still clustered first, the same is not
true for soft emissions anymore.
Thus the splitting structure within the anti-$k_t$ algorithm must be
constructed via the \kt splitting algorithm~\cite{Cacciari:2011ma}.

Just like QCD matrix elements, the \kt splitting scales provide a
unified view of initial- and final-state radiation. Through the combination of the
distance to the beams and the relative distance of objects to each other,
the $\sqrt{\di}$ distributions contain information about both the $p_\mathrm{T}$ spectra and
the substructure of jets.

\subsection{Existing predictions and measurements}

The \kt splittings and related distributions have attracted the attention of theorists, in $\Wln$ and similar final states.
They can be resummed analytically at next-to-leading-logarithm accuracy
as demonstrated for the example of jet production by QCD processes
in hadron collisions in Refs.~\cite{Banfi:2004nk,Banfi:2010xy}.
The ratio observable $y_{23}$ defined by the authors is closely
related to the ratio observables $\sqrt{d_{k+1}/d_k}$ in this
analysis. Other theoretical studies may be found in Refs.~\cite{mlm,Hoeche:2009rj}.

Experimentally, these kinds of observables were measured
at LEP~\cite{Abreu:1996na,Pfeifenschneider:1999rz,Heister:2003aj}
using the $e^+e^-$ (Durham) \kt algorithm.
Their theoretical features (resummability) were used in
Refs.~\cite{Dissertori:2009ik,Frederix:2010ne} to determine $\alpha_s$ with high precision.
Related observables were also measured at HERA~\cite{Adloff:1998ni,Tobien:1999rh,Chekanov:2002ux,Chekanov:2004kz}.

\section{Data analysis}
\label{sec:data}
\subsection{The ATLAS detector}

The ATLAS detector~\cite{Aad:2008zzm} at the LHC covers nearly the entire solid angle around the collision point.
It consists of an inner tracking detector surrounded by a thin superconducting solenoid, electromagnetic
and hadronic calorimeters, and a muon spectrometer incorporating three large superconducting
toroid magnets.

The inner-detector system is immersed in a 2~T axial magnetic field and provides charged particle
tracking in the range $|\eta| < 2.5$~\footnote{ATLAS uses a right-handed coordinate system with its origin at the nominal interaction point (IP) in the centre of the detector and the $z$-axis along the beam pipe. The $x$-axis points from the IP to the centre of the LHC ring, and the $y$-axis points upward. Cylindrical coordinates $(r,\phi)$ are used in the transverse plane, $\phi$ being the azimuthal angle around the beam pipe. The pseudorapidity is defined in terms of the angle $\theta$ as $\eta=-\ln\tan(\theta/2)$.}.
The high-granularity silicon pixel detector covers the vertex region
and typically provides three measurements per track.
It is followed by the silicon microstrip tracker which usually provides four two-dimensional measurement
points per track. These silicon detectors are complemented by the transition radiation tracker, which
contributes to track reconstruction up to $|\eta| = 2.0$. The transition radiation tracker also provides
electron identification information based on the fraction of hits (typically 30 in total) above a higher energy-deposit threshold corresponding to transition radiation.

The calorimeter system covers the pseudorapidity range $|\eta|< 4.9$. Within the region $|\eta|< 3.2$,
electromagnetic calorimetry is provided by barrel and endcap high-granu\-larity lead/liquid-argon (LAr)
calorimeters, with an additional thin LAr presampler covering $|\eta| < 1.8$ to correct for
energy loss in material upstream of the calorimeter. Hadronic calorimetry is provided by a steel/scintillator-tile
calorimeter, segmented radially into three barrel structures within $|\eta| < 1.7$, and two copper/LAr hadronic endcap
calorimeters. The solid angle coverage is completed with forward copper/LAr and tungsten/LAr calorimeter modules
optimised for electromagnetic and hadronic measurements respectively.

The muon spectrometer comprises separate trigger and high-precision tracking chambers measuring
the deflection of muons in a magnetic field generated by superconducting air-core toroids. The precision
chamber system covers the region $|\eta| < 2.7$ with three layers of monitored drift tubes, complemented by
cathode strip chambers in the forward region, where the background is highest. The muon trigger system
covers the range $|\eta| < 2.4$ with resistive plate chambers in the barrel, and thin gap chambers in the endcap
regions.

A three-level trigger system is used to select interesting events~\cite{Aad:2012xs}. The Level-1 trigger is implemented
in hardware and uses a subset of detector information to reduce the event rate to a design value of at most 75~kHz. This
is followed by two software-based trigger levels which together reduce the event rate to about 200~Hz.

\subsection{Event selection}
The selection of $W$ events is based on the criteria described in Refs.~\cite{Aad:2012en,WZinc} and summarised briefly below.

\subsubsection{Data sample and trigger}
The entire 2010 data sample at $\sqrt{s}=7$~TeV was used, corresponding to an integrated luminosity of approximately 36 $\rm{pb}^{-1}$. The 2010 data sample was chosen due to the low pileup conditions during data taking, where the mean number of interactions per bunch crossing was at most $2.3$ during that period. In the $\Wmn$ analysis, the first few $\rm{pb}^{-1}$ were excluded to restrict to a data sample of events recor\-ded with a uniform trigger configuration and optimal detector performance.

Single-lepton triggers were used to retain $\Wln$ candidate events. For the electron channel a trigger threshold of $14\gev$ for early data-taking periods and  $15\gev$ for later data-taking periods was applied. For the muon channel a trigger threshold of $13\gev$ was applied. All relevant detector components were required to be fully operational during the data taking. Events with at least one reconstructed interaction vertex within 200~mm of the interaction point in the $z$ direction and having at least three associated tracks were considered. The number of reconstructed vertices reflects the pileup conditions and, in both channels, was used to reweight the MC simulation to improve its modelling of the pileup conditions observed in data. The number of reconstructed vertices was also used to estimate the uncertainty due to possible mismodelling of the pileup.

\subsubsection{Electron selection}
Clusters formed from energy depositions in the electromagnetic calorimeter were required to have matched tracks, with the further requirement that the cluster shapes are consistent with electromagnetic showers initiated by electrons. On top of the tight identification criteria, a calori\-meter-based isolation requirement for the electron was applied to further reduce the multi-jet background. Additional requirements were applied to remove electrons falling into calorimeter regions with non-operational LAr readout. The kinematic requirements on the electron candidates included a transverse momentum requirement $p_\mathrm{T}^\ell>20$~GeV and pseudorapidity $|\eta^\ell|<2.47$ with removal of the transition region $1.37<|\eta^\ell|<1.52$ between the calorimeter modules. Exactly one of these selected electrons was required for the $\Wen$ selection. In constructing the \kt cluster sequence, clusters of calo\-rimeter cells included in a reconstructed jet within $\Delta R=0.3$ of the electron candidate were removed from the input configuration.

\subsubsection{Muon selection}
Muon candidates were required to have tracks reconstruc\-ted in both the muon spectrometer and inner detector, with $p_\mathrm{T}^\ell$ above 20~GeV and pseudorapidity $|\eta^\ell|<2.4$. Requirements on the number of hits used to reconstruct the track in the inner detector were applied, and the muon's point of closest approach to the primary vertex was required to be displaced in $z$ by less than 10~mm. Track-based isolation requirements were also imposed on the reconstructed muon. At least one muon was required for the $\Wmn$ selection. 
 To retain consistency with the acceptance in the electron channel, when constructing the \kt cluster sequence, clusters of calorimeter cells falling close to the muon candidate were removed from the input configuration as in the electron selection.

\subsubsection{Selection of $W$ candidate events and construction of observables}

\vspace*{-2mm}
The $\Wln$ event selection required that the magnitude of the missing transverse momentum, $\met$\cite{EtMiss}, be greater than 25~GeV. The reconstructed transverse mass obtained from the lepton transverse momentum $\vec{p}_\mathrm{T}^{\,\ell}$ and $\vec{E}_{\mathrm{T}}^{\mathrm{miss}}$~vectors was required to fulfil \\$m_{\mathrm{T}}^W=\sqrt{2(p_{\rm{T}}^\ell\met-\vec{p}_{\rm{T}}^{\,\ell}\cdot\vec{E}_{\mathrm{T}}^{\mathrm{miss}})}>40$~GeV. No requirements were made with respect to the number of reconstructed jets in the event.

The observables defined in Sect.~\ref{sec:variables} were constructed using calorimeter energy clusters within a pseudorapidity range of $|\eta^{\rm{cl}}|<4.9$. The clusters were seeded by calorimeter cells with energies at least 4$\sigma$ above the noise level. The seeds were then iteratively extended by including all neighbouring cells with energies at least 2$\sigma$ above the noise level. The cell clustering was finalised by the inclusion of the outer perimeter cells around the cluster. The so-called topological clusters that resulted were calibrated to the hadronic energy scale~\cite{LCCali1,LCCali2}, by applying weights to account for calorimeter non-compensation, energy lost upstream of the calorimeters and noise threshold effects.

\vspace*{-2mm}
\subsection{Background treatment}
\vspace*{-2mm}
The contributions of electroweak backgrounds ($Z\to \ell \ell$, $W\to\tau\nu$ and diboson production), as well as $t\overline{t}$ and single-top-quark production, to both channels were estimated using the MC simulation. The absolute normalisation was derived using the total theoretical cross sections and corrected using the acceptance and efficiency losses of the event selection. The shape and normalisation of the distributions of various observables for the multi-jet background were determined using data-driven methods in both analysis channels. For the $\Wen$ selection, the background shape was obtained from data by reversing certain calorimeter-based electron identification criteria to produce a multi-jet-enriched sample. Similarly, to estimate the multi-jet contribution to $\Wmn$, the background shape was obtained from data by inverting the requirements on the muon transverse impact parameter and its significance.
These multi-jet enriched samples provided the shapes of the distributions of multi-jet background observables. The normalisation of the multi-jet background was determined by fitting a linear combination of the multi-jet and leptonic  \met~ shapes to the observed \met~ distribution, following the procedures described in Refs.~\cite{Aad:2012en,WZinc}. The total background was thus estimated to be 5\% of the signal for the $\Wen$ analysis, with the largest contribution arising from multi-jet production.  For the $\Wmn$ analysis, the total background is 9\% of the signal and is dominated by the $Z\to \ell \ell$ process.  At large splitting scales, top quark pair production becomes the dominant contribution in both channels.

\section{Monte Carlo simulations}
\label{sec:mc}
All detector-level studies and the extraction of particle-level distributions involved two signal MC generators, \alpgen and \sherpa.
\textsc{Alpgen} v2.13~\cite{Mangano:2002ea}, a matrix-element (ME) generator, was interfaced to \textsc{Herwig}\linebreak v6.510~\cite{herwig} for parton showering (PS) and hadronisation, and to \textsc{Jimmy} v4.31~\cite{jimmy} for multiple parton interactions. The MLM~\cite{mlm} matching scheme was used to combine $W$-boson production samples having up to five partons with the parton shower, with the matching scale set at 20~GeV. 
\sherpa v1.3.1~\cite{Sherpa} was used to generate an alternative signal sample of events with $W$ + jets, using a ME+PS merging approach~\cite{Hoeche:2009rj} to prevent double counting from the parton shower, and extending the original CKKW method~\cite{ckkw} by taking into account truncated shower emissions. Up to five partons were generated in the ME and the matching scale was set to 30~GeV. 

The single-top-quark background events were generated at next-to-leading-order (NLO) accuracy using the \mcatnlo v3.3.1~\cite{Frixione:2002ik} generator.  \mcatnlo was interfaced to \textsc{Herwig} and \textsc{Jimmy}.
The \textsc{Powheg} v1.01~\cite{Frixione:2007vw} generator, interfaced to \textsc{Pythia6} v6.421~\cite{Sjostrand:2006za}, was used to simulate the $t\bar{t}$ background. The background from diboson production was generated using \textsc{Herwig}. Backgrounds from inclusive $Z$ production were simulated using \textsc{Pythia6}.

Three sets of parton density functions (PDFs) were used in these MC samples: CTEQ6L1~\cite{Pumplin:2002vw} for the \textsc{Alpgen} samples and the parton showering and underlying event in the \textsc{Powheg} samples interfaced to \textsc{Pythia6}; MRST 2007 LO$^\ast$~\cite{Sherstnev:2007nd} for \textsc{Pythia6} and \textsc{Herwig}; and CTEQ6.6M~\cite{Nadolsky:2008zw} for \mcatnlo, \sherpa, and the NLO matrix element calculations in \textsc{Powheg}. The underlying event tunes were AUET1~\cite{auet} for the \textsc{Herwig}, \textsc{Alpgen}, and \mcatnlo samples, and AMBT1~\cite{ambt1} for the \textsc{Pythia6} and \textsc{Powheg} samples. The samples generated with \sherpa used the default underlying event tune.

Each generated event was passed through the standard ATLAS detector simulation \cite{detsim}, based on \textsc{Geant4}~\cite{geant}. The MC events were reconstructed and analysed using the same software chain as applied to the data. The resulting MC predictions for the samples were normalised to their respective theoretical cross sections calculated at NLO~\cite{Aad:2012en}, with the exception of the $W$ and $Z$ samples which were normalised to NNLO~\cite{Melnikov:2006kv}, and the multi-jet background which was normalised to a value extracted from the data as is described in Sect.~\ref{sec:data}.

At the particle level, some additional $W$+jets NLO MC generators were compared to the final results. The \textsc{Powheg}~\cite{Frixione:2007vw,Alioli:2008gx} samples were matched to \textsc{Pythia6} v6.425 or \textsc{Pythia8} v8.165~\cite{Sjostrand:2007gs} for parton showering and hadronisation, while another sample was generated with \mcatnlo v4.06~\cite{Frixione:2002ik} using \textsc{Herwig} v6.520.2. The \sherpa \textsc{Menlops} sample used \sherpa v1.4.1 with its built-in \textsc{Menlops} method~\cite{Hoche:2010kg}, allowing an NLO+PS matched sample for inclusive $W$ production~\cite{Hoeche:2011fd} to be merged with LO matrix elements for a $W$ boson and up to five partons using a matching scale at 20~GeV. All these NLO samples were generated with the CT10 PDF set~\cite{Lai:2010vv}.

The \mcatnlo, \textsc{Powheg} and \alpgen samples were supplemented with a simulation of QED final-state radiation using \textsc{Photos} v2.15.4~\cite{Golonka:2005pn} and tau decays using \textsc{Tauola} v27feb06~\cite{Jadach:1990mz}. The \sherpa samples included QED final-state radiation in a different resummation approach~\cite{Schonherr:2008av} and a built-in tau decay algorithm.

\section{Detector-level comparisons of Monte Carlo to data}
\label{sec:DetLevelDists}
  \begin{figure*}[htbp]  
    \begin{center}
      \includegraphics[width=0.4\textwidth]{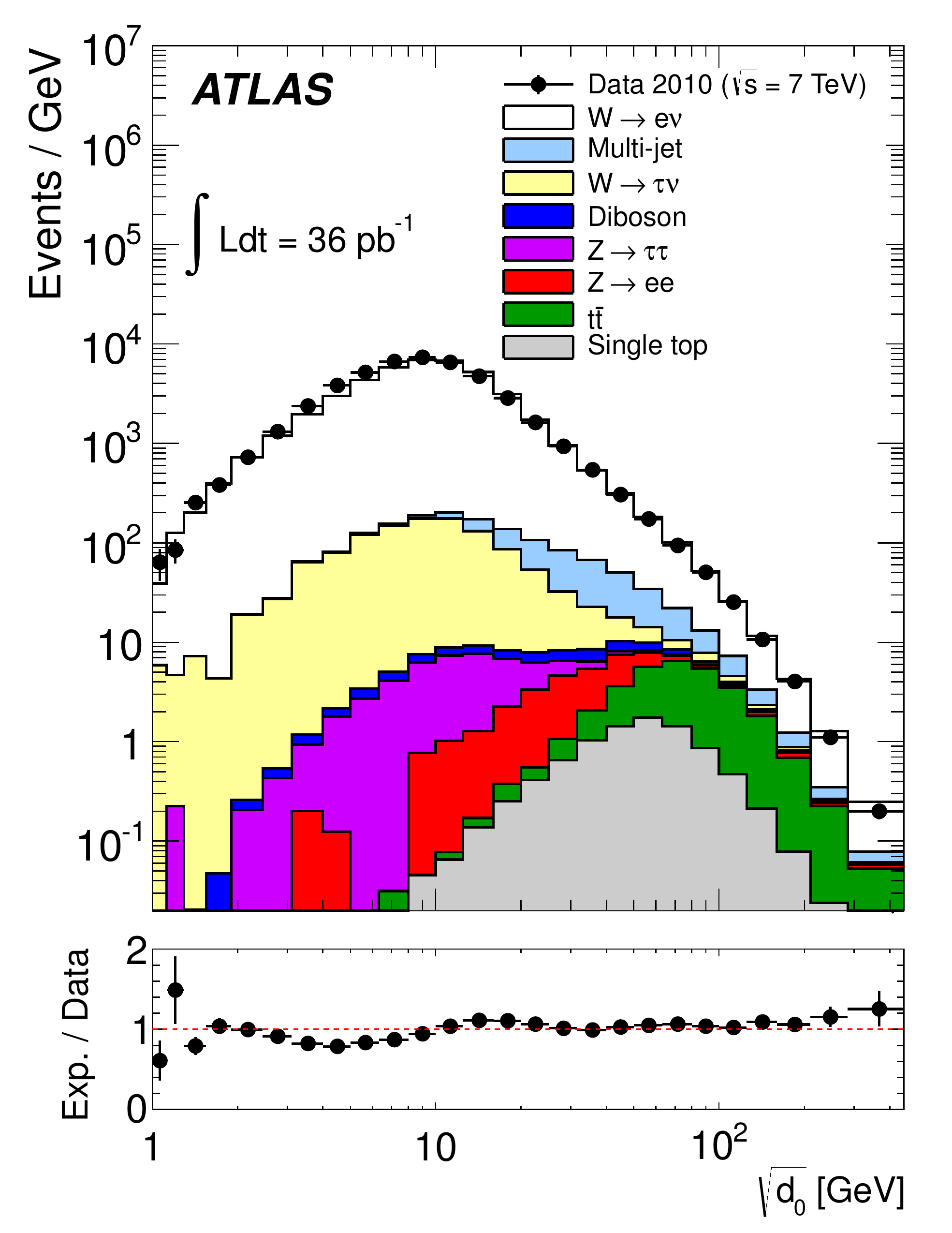}
      \includegraphics[width=0.4\textwidth]{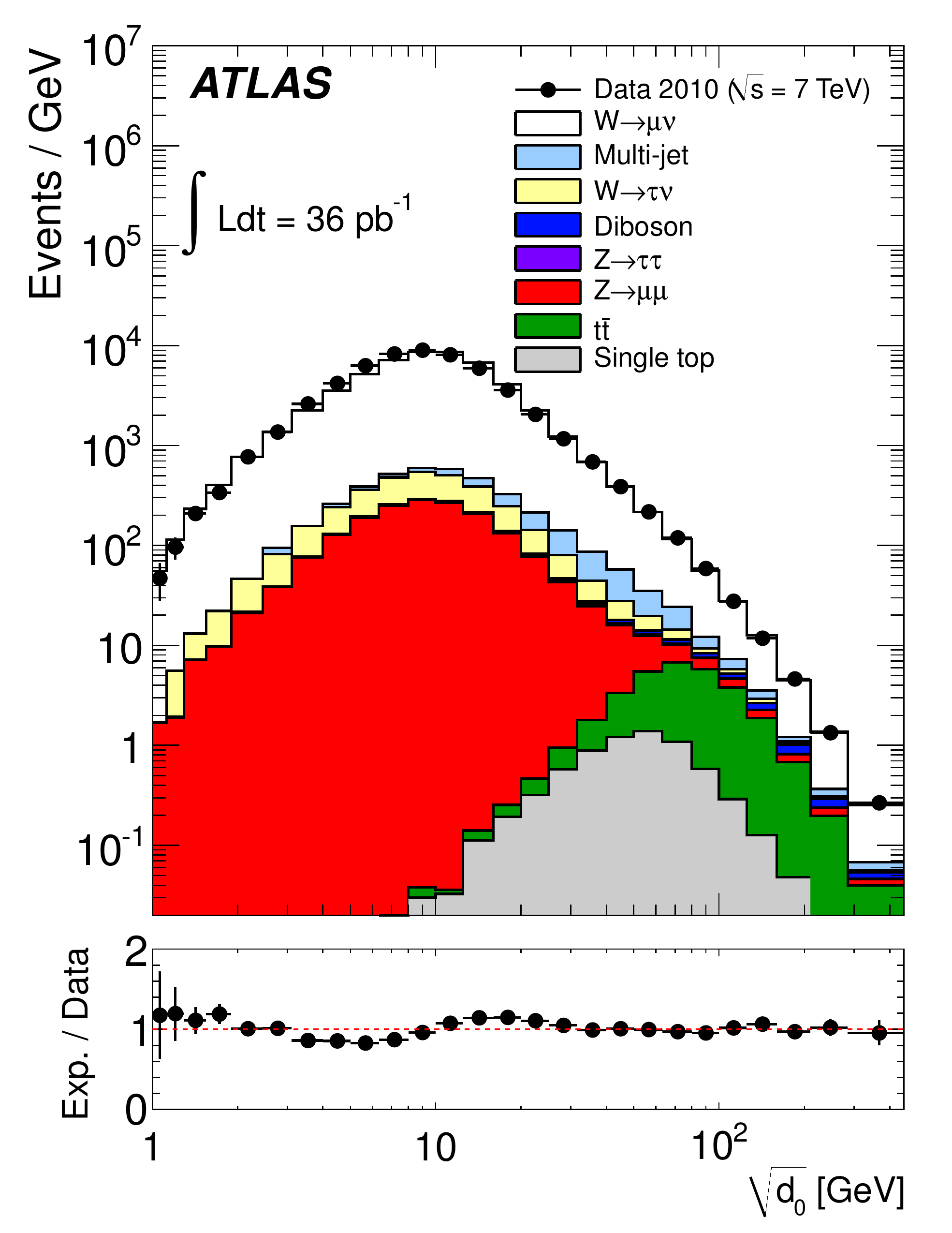}
      \caption{Uncorrected splitting scale $\sqrt{d_{0}}$ for events passing the $W\to e\nu$ (left) and $W\to \mu\nu$ (right) selection requirements. The distributions from the data (markers) are compared with the predicted signal from the MC simulation, provided by \alpgen and normalised to the NNLO prediction. In addition, physics backgrounds, also shown, have been added in proportion to the predictions from the MC simulation. The ratio between the expectation and the data is shown in the lower plot. The error bars shown on the data are statistical only.}
      \label{fig:det_d0}
    \end{center}
  \end{figure*}

  \begin{figure*}[htbp]  
    \begin{center}
      \includegraphics[width=0.4\textwidth]{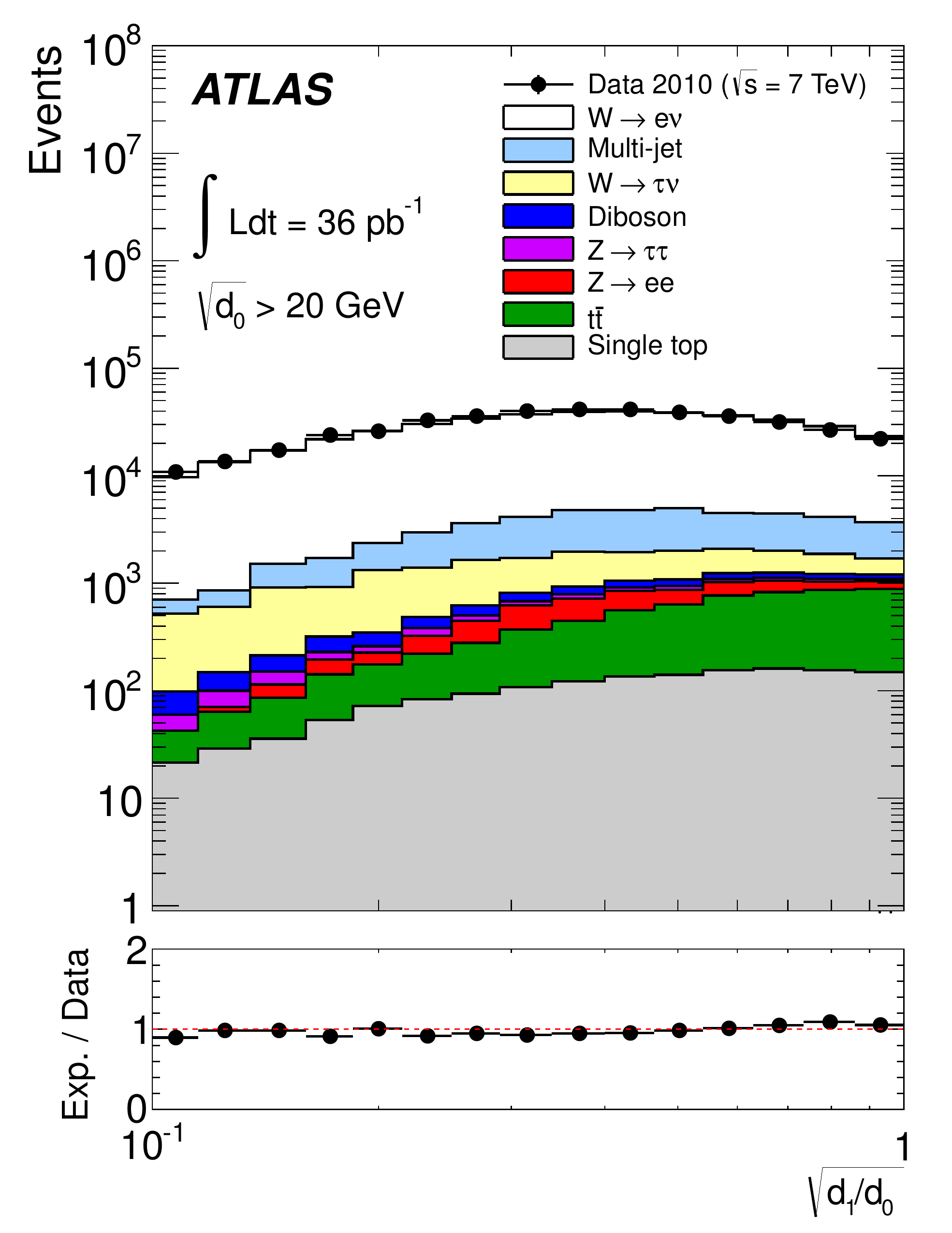}
      \includegraphics[width=0.4\textwidth]{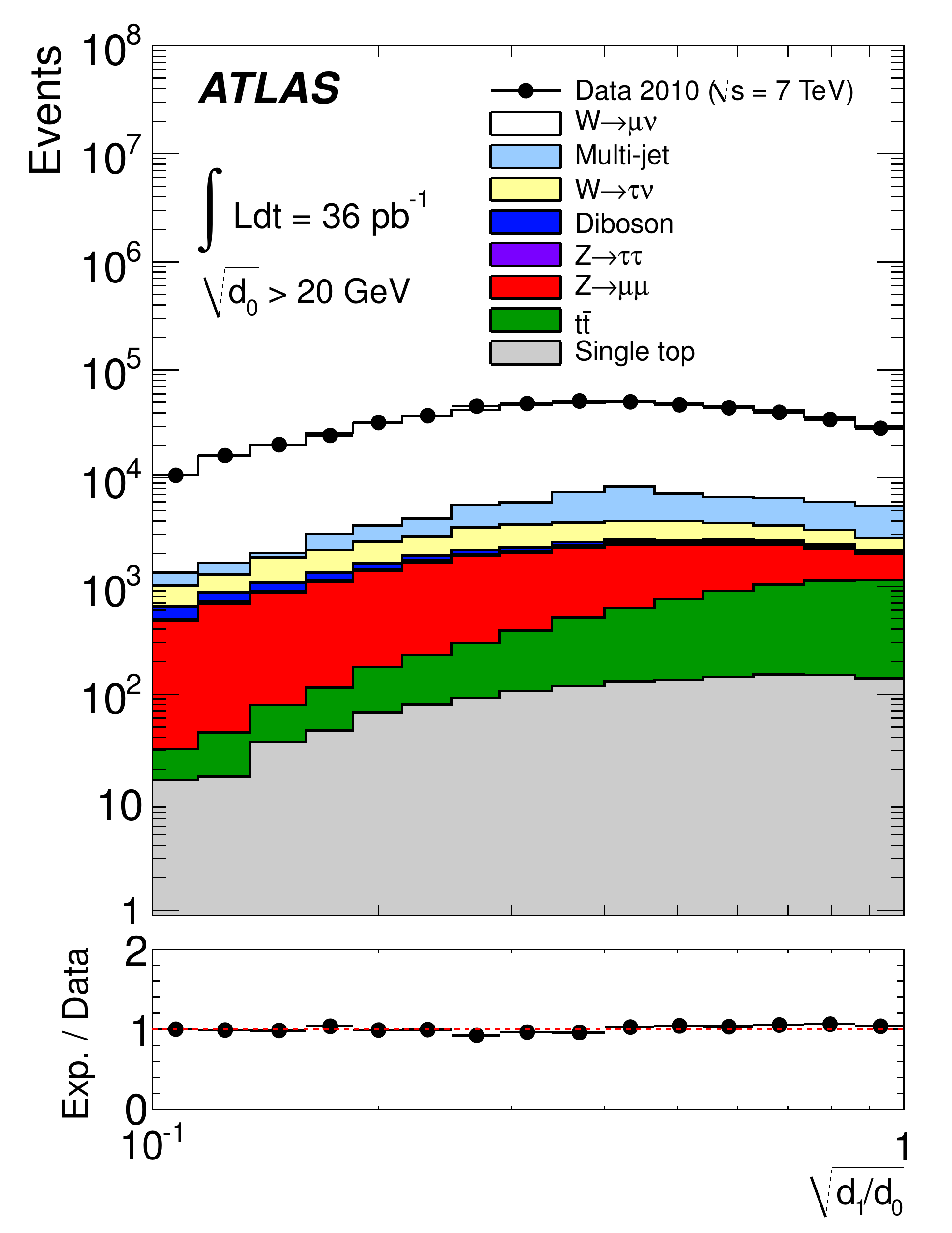} 
      \caption{Uncorrected ratio $\sqrt{d_{1}/d_{0}}$ for events passing the $W\to e\nu$ (left) and $W\to \mu\nu$ (right) selection requirements.  The distributions from the data (markers) are compared with the predicted signal from the MC simulation, provided by \alpgen and normalised to the NNLO prediction. In addition, physics backgrounds, also shown, have been added in proportion to the predictions from the MC simulation. The ratio between the expectation and the data is shown in the lower plot. The error bars shown on the data are statistical only.}
      \label{fig:det_y1}
    \end{center}
  \end{figure*}

  The observed and expected detector-level distributions for $\sqrt{d_{0}}$ in the electron and muon channels are shown in Fig.~\ref{fig:det_d0}, where the MC signal predictions are provided by \alpgen normalised to NNLO predictions~\cite{Melnikov:2006kv}.
The $W$-boson kinematic distributions are shown in detail in Refs.~\cite{Aad:2012en,WZinc}.
The corresponding plots for $\sqrt{d_{1}}$, $\sqrt{d_{2}}$ and $\sqrt{d_{3}}$ can be found in Appendix~\ref{sec:adddists}. Figure \ref{fig:det_y1} shows the ratio of the second-hardest to the hardest splitting scale in each event. Again, the sub-leading ratio distributions at detector level are displayed in Appendix~\ref{sec:adddists}.
For the hardest clustering in the event, $\sqrt{d_0}$, generally good agreement between the \alpgen MC predictions and the data is observed. The agreement is similar for both the electron and the muon channels.

\section{Particle-level extraction}
\label{sec:particlelevel}

\subsection{Corrections for detector effects}
\label{sec:unfolded}
After subtraction of backgrounds, the detector level distributions were corrected (``un\-fol\-ded'') to the final-state particle level separately for the two channels, taking into account the effects of pileup and detector response. The unfolding was performed with the RooUnfold~\cite{RooUnfold} package, using a Baye\-sian algorithm\cite{D'Agostini:1994zf}, in which Bayes’ theorem was used to derive the particle-level distributions from the detector-level distributions, over three iterations. The input for the algorithm at particle and detector level was taken from the \alpgen sample as a default. Both the MC simulation and data-driven methods were used to demonstrate that this iterative Bayesian method was able to recover the corresponding particle-level distributions. 

The selection requirements applied to the event at the particle level are:

\begin{itemize}
\item[$\bullet$] $p_\mathrm{T}^\ell > $ 20~GeV ($\ell=\text{electron }e\text{ or muon }\mu$)
\item[$\bullet$] $|\eta^e| < $ 2.47 excluding  $1.37<|\eta^e|<1.52$
\item[$\bullet$] $|\eta^\mu| < $ 2.4
\item[$\bullet$] $p_{\rm{T},\text{lead}}^{\nu} > $ 25~GeV ($\nu_\text{lead}=\text{highest-$p_{\rm{T}}$ neutrino in event}$)
\item[$\bullet$] $m_{\mathrm{T}}^W > $ 40~GeV
\end{itemize}

Only events with exactly one lepton passing the requirements were taken into account. Leptons were defined to include all photon radiation within a cone of $\Delta R=0.1$ around the final-state lepton as suggested in Ref.~\cite{Butterworth:2010ym}. All lepton requirements were calculated from these combined objects. The observables defined in Sect.~\ref{sec:variables} were constructed using all stable particles within a pseudorapidity range of $|\eta^{\rm{cl}}|<4.9$ with lifetime greater than 10~ps, excluding the lepton and neutrino originating from the $W$ boson decay.

\subsection{Weighted combination}
\label{sec:weightedcombo}

To reduce the impact of imperfect MC modelling of pileup effects, whilst optimising the statistical power available, two different event samples were defined and utilised as follows.

\begin{itemize}
\item ``Low-pileup sample'': exactly one reconstructed vertex was required in data. The response matrices used to unfold the data and the background templates were also constructed from events where exactly one reconstructed vertex was required. 
\item ``High-pileup sample'': as above, with the difference that the number of reconstructed vertices was required to be greater than one. 
\end{itemize}

At large $\sqrt{d_k}$, the statistical uncertainty of the high-pileup sample is smaller than that in the low-pileup sample. However, at small $\sqrt{d_k}$, the systematic pileup uncertainty of the low-pileup sample is smaller than that in the high-pileup sample. To minimise the overall uncertainty on the measurement, the distributions were combined as follows.
For each bin of the final distribution, the best estimate $N$ was calculated from the bin contents $N_1$, $N_2$ of the distributions in the low-pileup and high-pileup samples respectively, as 

\begin{equation}
\label{eq_comb_mean}
N=\frac{N_1\cdot{W_1} + N_2\cdot{W_2}}{{W_1}+{W_2}}.
\end{equation}

\noindent The weights $W_i$ for each sample were constructed from the inverse of the sum in quadrature of the statistical and pileup uncertainties on the low-pileup and the high-pileup samples. The evaluation of the pileup uncertainty on each sample is described in detail in Sect.~\ref{sec:unc}. The statistical uncertainty of the final distribution was calculated assuming no correlation between the two samples.

\section{Systematic uncertainties}
\label{sec:unc}
To evaluate the impact of a particular source of systematic uncertainty at the particle level, the observable considered was varied within its uncertainty, the response matrix was recalculated taking this variation into account, and the new response matrix was used to unfold the data. The fractional shift in the resulting unfolded data from nominal was interpreted as the systematic uncertainty due to that particular effect. The separate sources of uncertainty are described in the following.

The relative systematic uncertainty on the energy scale of the topological clusters was evaluated from a combination of MC studies and single-pion response measurements~\cite{EtMiss} to be $1 \pm a \times (1 + b / p_{\rm{T}}^{\rm{cl}})$ where $p_{\rm{T}}^{\rm{cl}}$ represents the transverse momentum of each cluster. The constants $a$ and $b$ were determined to be $a = 3~(10)\%$ when $|\eta^{\rm{cl}}|<3.2$ ($|\eta^{\rm{cl}}|>3.2$), and $b=1.2~$GeV. A shift of the cluster energy results in a shift of the distributions to higher or lower values. The uncertainty due to the cluster energy scale was thus evaluated separately for the low-pileup and high-pileup distributions and combined in a weighted linear sum. The uncertainty  ranges from 5$\%$ to 55$\%$ for the splitting scales $\sqrt{d_{k}}$ and from 2$\%$ to 85$\%$ for the $\sqrt{d_{k+1}/d_{k}}$ ratio distributions.  

The lepton trigger, identification and reconstruction efficiencies as well as the lepton energy scale and resolution were measured in data using $Z\to \ell\ell$ events via the tag-and-probe method, as described in Refs.~\cite{Aad:2012en,Aad:2011mk,WZinc}. The uncertainty is less than 3$\%$ for the splitting scales $\sqrt{d_{k}}$ and less than 1$\%$ for the $\sqrt{d_{k+1}/d_{k}}$ ratio distributions.

The systematic uncertainty due to possible MC mismodelling of pileup was evaluated separately on the low-pileup and high-pileup distributions. The impact of pileup mismodelling on the low-pileup sample was evaluated by varying the requirements on the $z$-displacement of the interaction vertex and the number of associated tracks. An additional uncertainty accounts for the possible mismodelling of contributions from adjacent
bunch-crossings. It was evaluated by comparing two different data-taking periods: one in which proton bunches were arranged in trains, and the other without bunch trains. 
The impact of pileup mismodelling on the high-pileup sample was evaluated as the fractional difference between the particle-level measurements for the low-pileup and the high-pileup events, with the statistical uncertainty subtracted in quadrature. The uncertainty ranges from 1$\%$ to 30$\%$ for the splitting scales $\sqrt{d_{k}}$ and is largest for small splitting scales. For the $\sqrt{d_{k+1}/d_{k}}$ ratio distributions the uncertainty ranges from 1$\%$ to 15$\%$. 

The uncertainty inherent in the unfolding procedure itself was estimated by reweighting the response matrix in the unfolding such that \alpgen would accurately model the distribution under consideration as measured from data at reconstruction level. A second variation was performed by creating a response matrix from \textsc{Sherpa}.  The larger effect, per bin, obtained from these two estimates of the systematic uncertainty was taken as the systematic uncertainty due to unfolding. The uncertainty ranges between 5$\%$ and 55$\%$ for the splitting scales $\sqrt{d_{k}}$, being largest for small values of $\sqrt{d_{k}}$ and in the vicinity of $\sqrt{d_{k}} \approx 15\gev$. For the $\sqrt{d_{k+1}/d_{k}}$ ratio distributions the uncertainty ranges between 1$\%$ and 35$\%$.

The systematic uncertainties on the electroweak and top-quark background normalisations were assigned using the theoretical uncertainty on the cross section of each process under consideration. The uncertainty on the multi-jet background normalisation was obtained by varying the methods used for extracting this value from data, as described in Refs.~\cite{Aad:2012en,WZinc}. An additional uncertainty was included on the shape of the multi-jet contribution, which was derived by comparing data-driven and simulation estimates of this background contribution. The uncertainty ranges from 0.5$\%$ to 15$\%$ for the splitting scales $\sqrt{d_{k}}$ and from 1$\%$ to 20$\%$ for the $\sqrt{d_{k+1}/d_{k}}$ ratio distributions. 

  \begin{figure*}[htbp]  
    \begin{center}
      \includegraphics[width=0.5\textwidth]{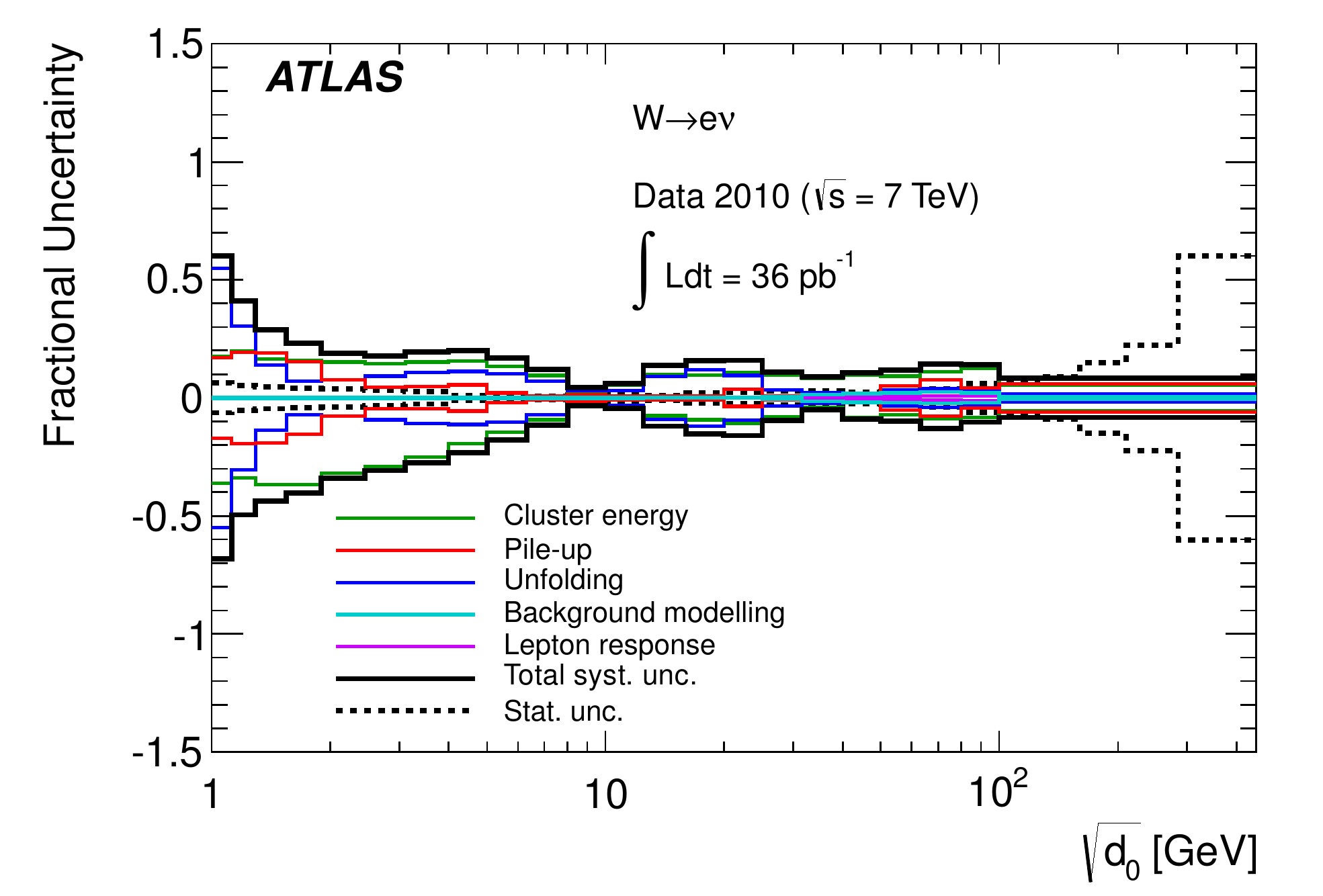}\nolinebreak
      \includegraphics[width=0.5\textwidth]{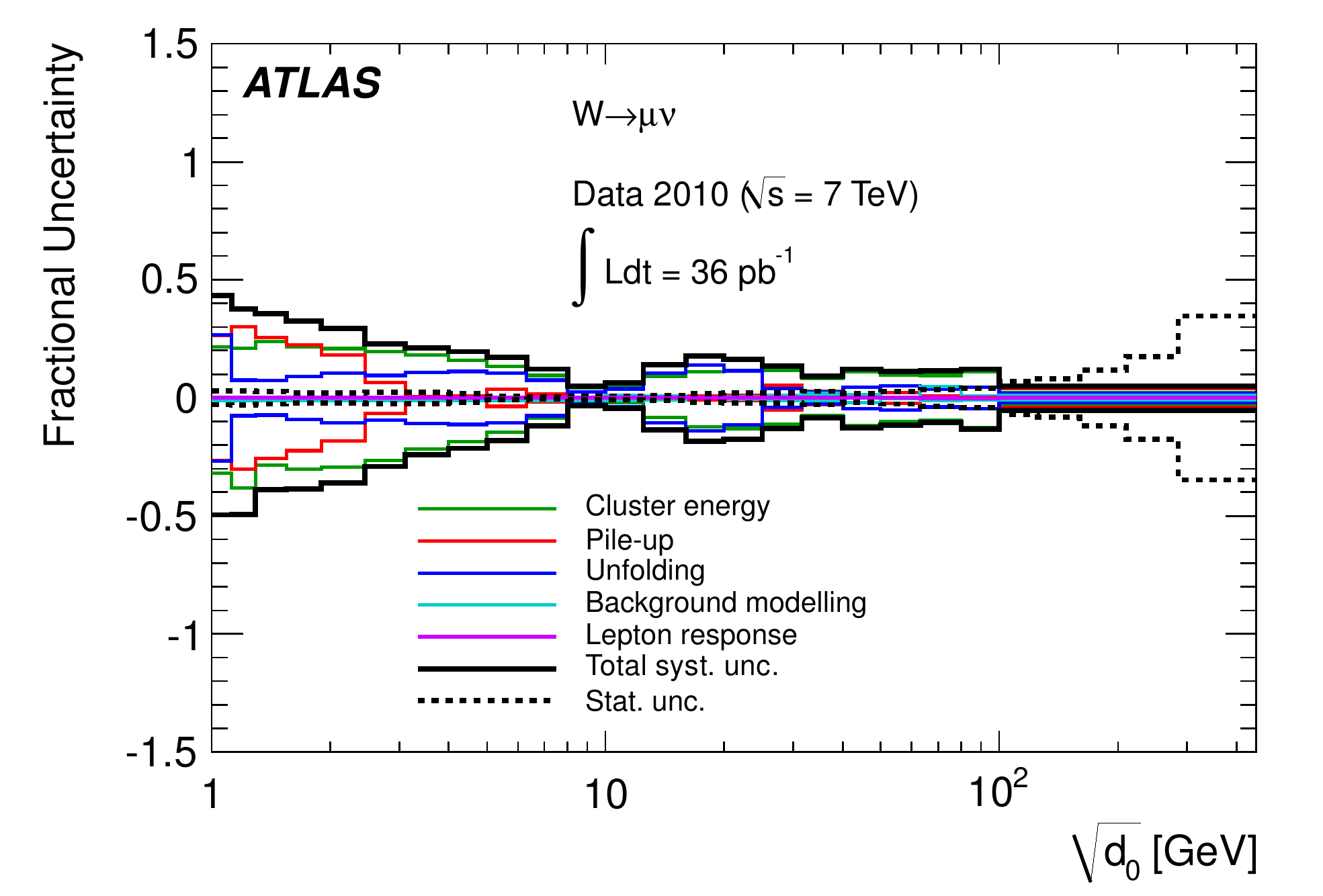}\\
      \includegraphics[width=0.5\textwidth]{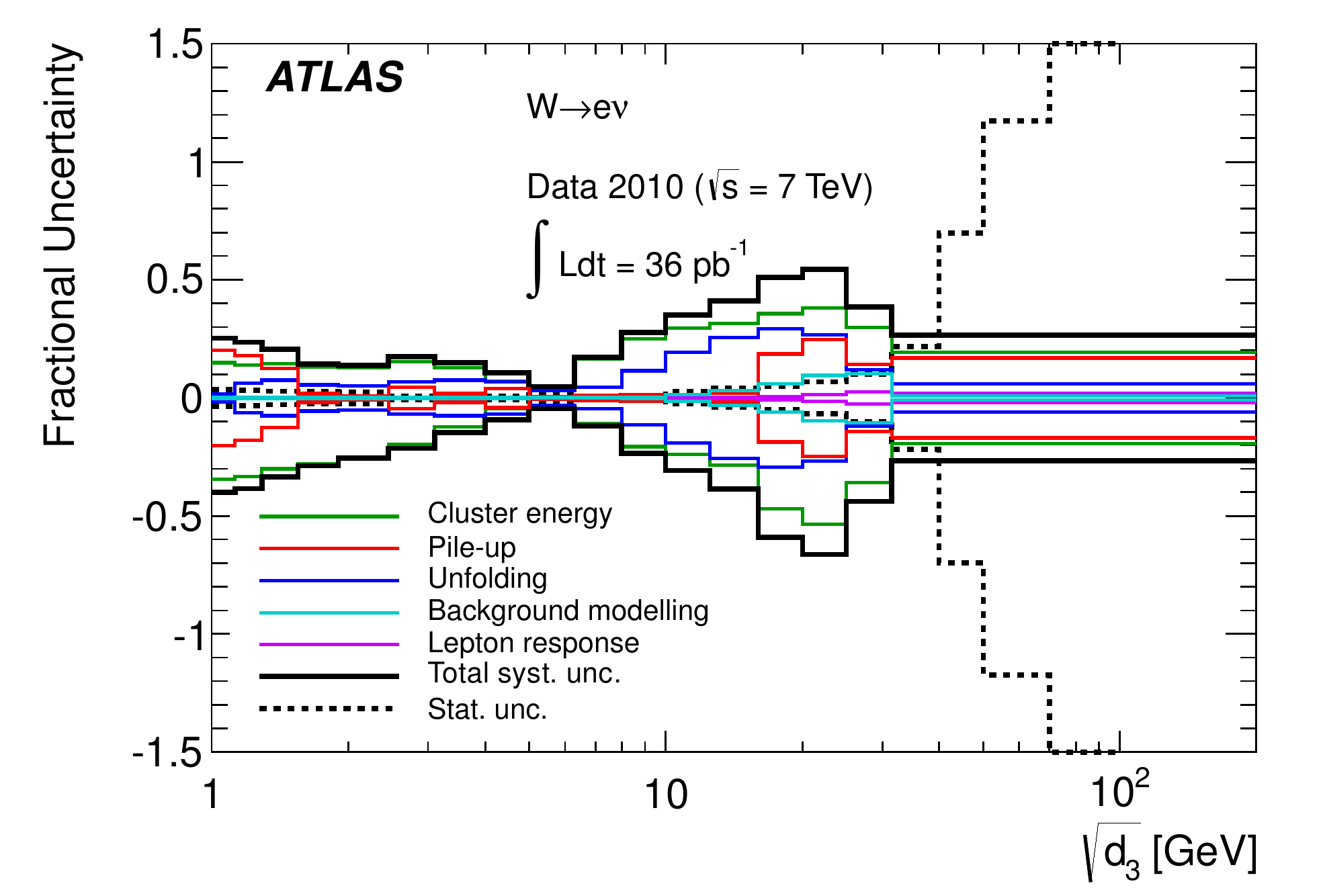}\nolinebreak
      \includegraphics[width=0.5\textwidth]{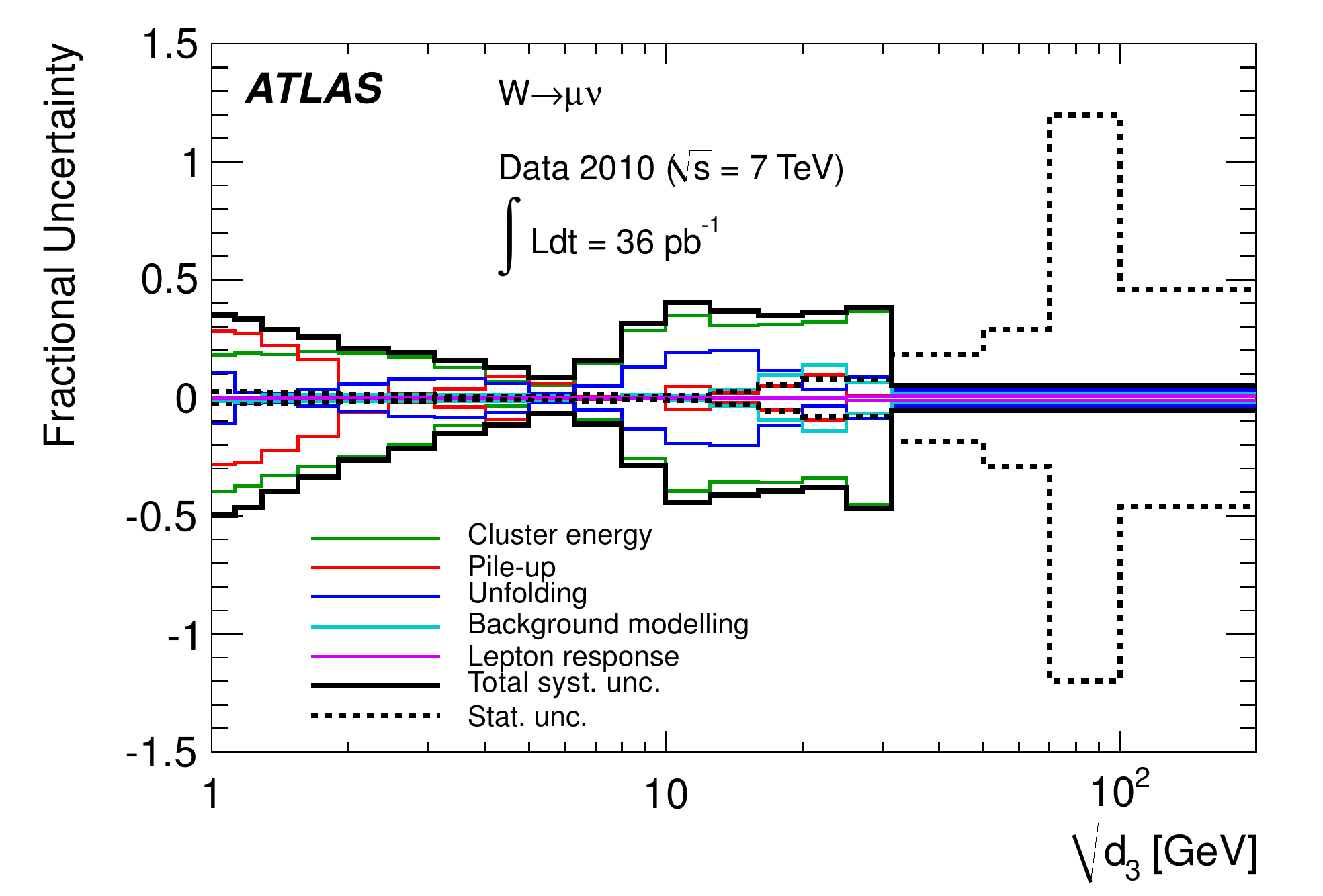}
      \caption{Summary of the systematic uncertainties on the measured particle-level distributions for $\sqrt{d_{0}}$ (top) and $\sqrt{d_{3}}$ (bottom) in the $W\to e\nu$ (left) and $W\to \mu\nu$ (right) channels.}
      \label{figure_overallunc_d03}
    \end{center}
  \end{figure*}
  \begin{figure*}[htbp]  
    \begin{center}
      \includegraphics[width=0.5\textwidth]{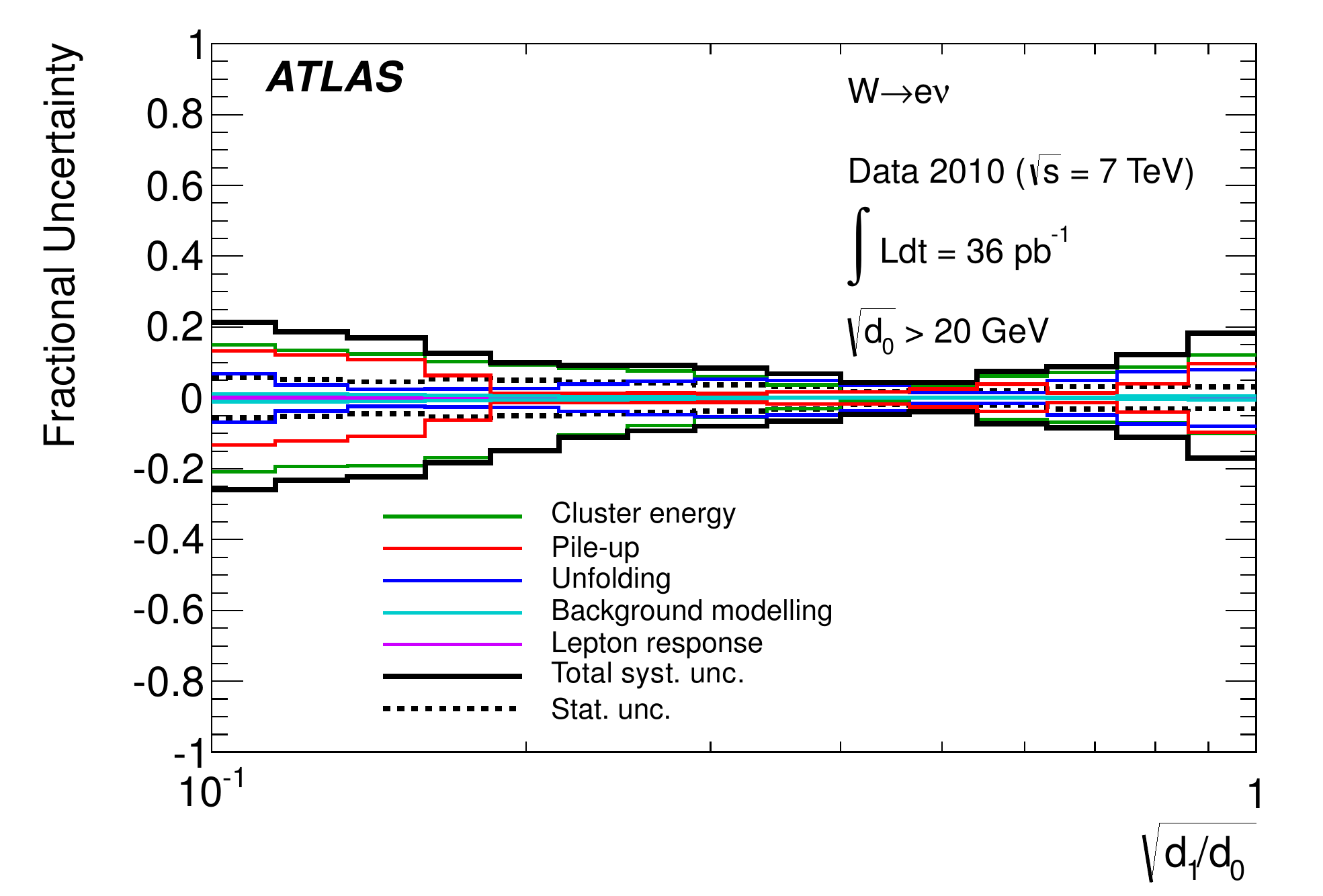}\nolinebreak
      \includegraphics[width=0.5\textwidth]{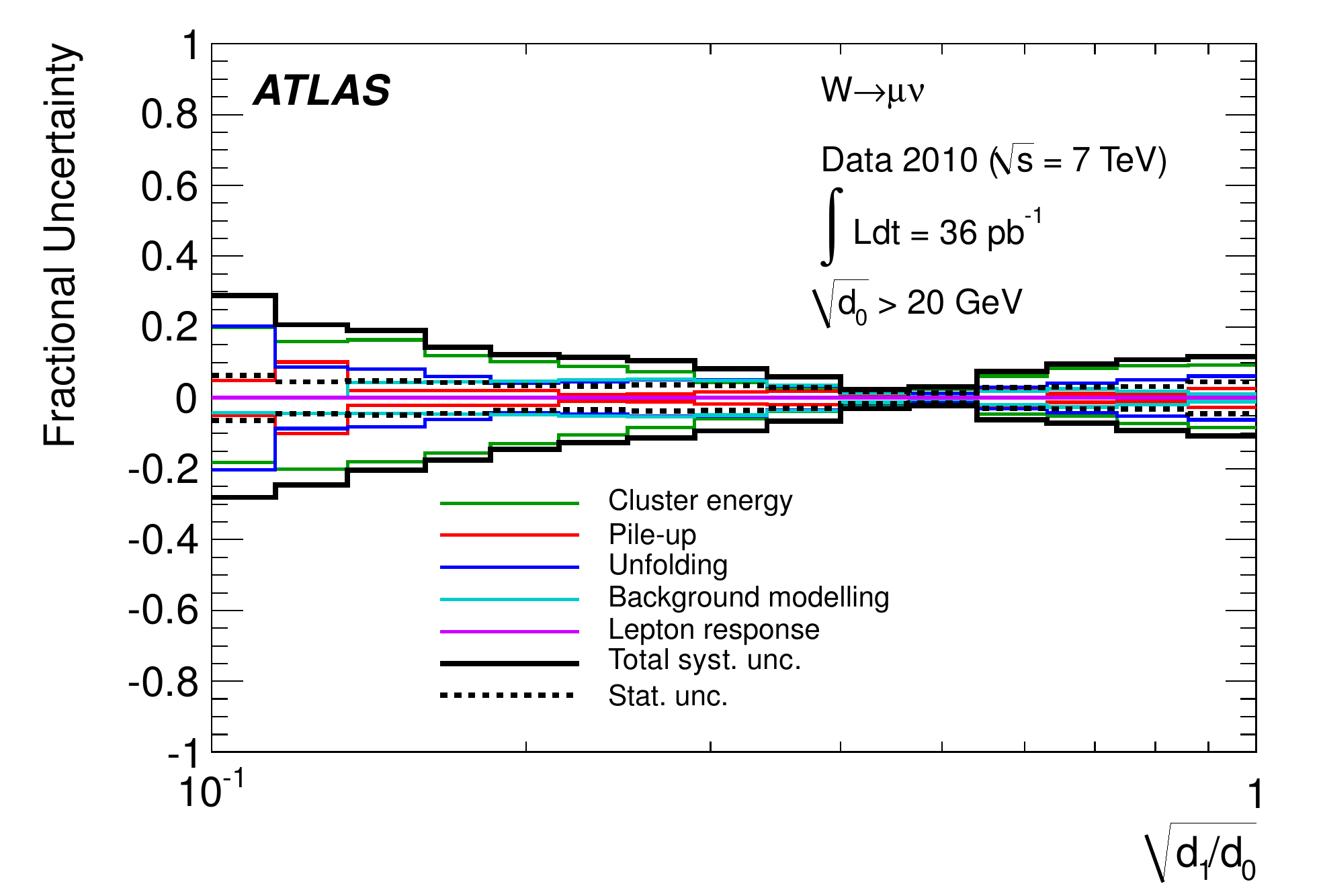}\\
      \includegraphics[width=0.5\textwidth]{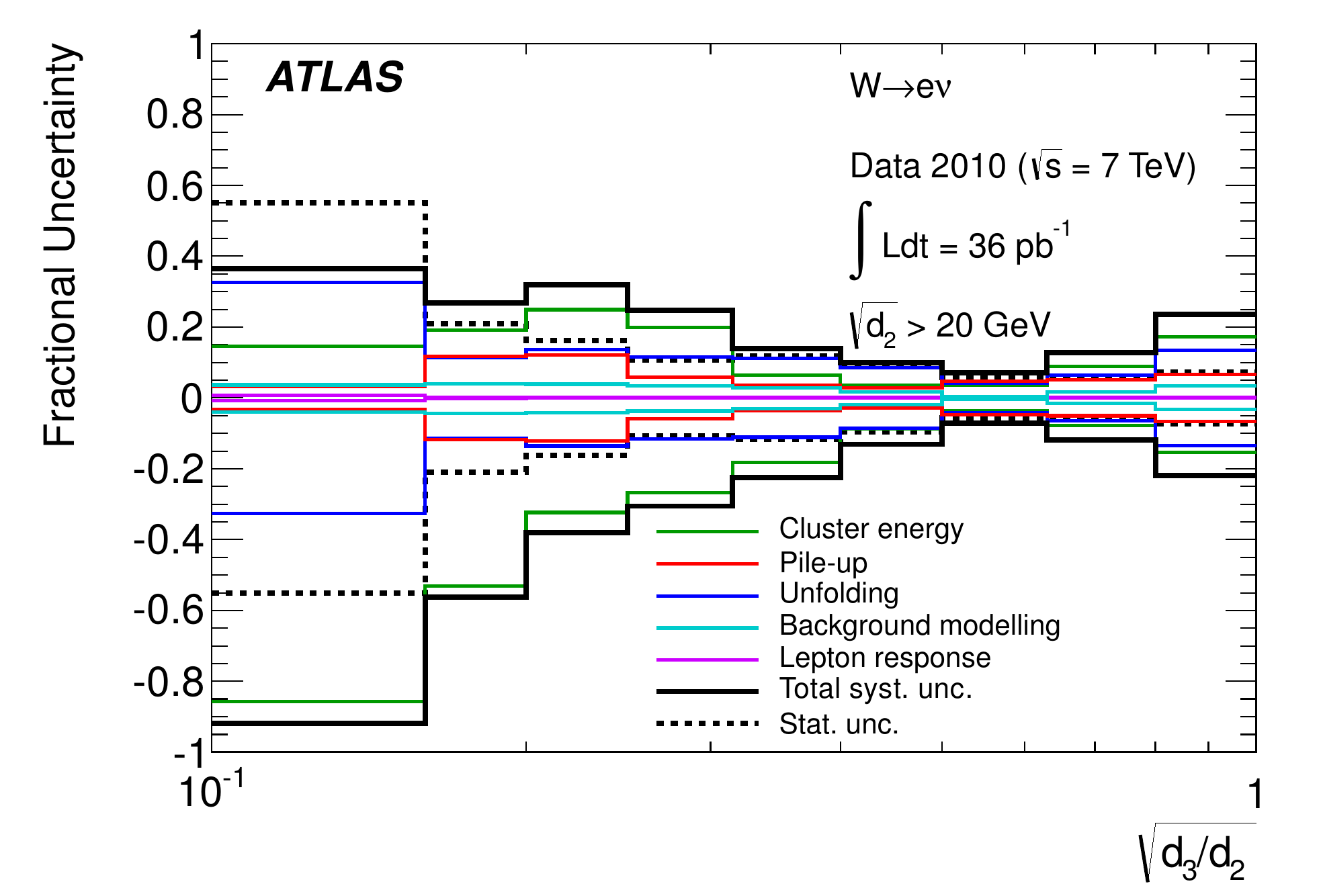}\nolinebreak
      \includegraphics[width=0.5\textwidth]{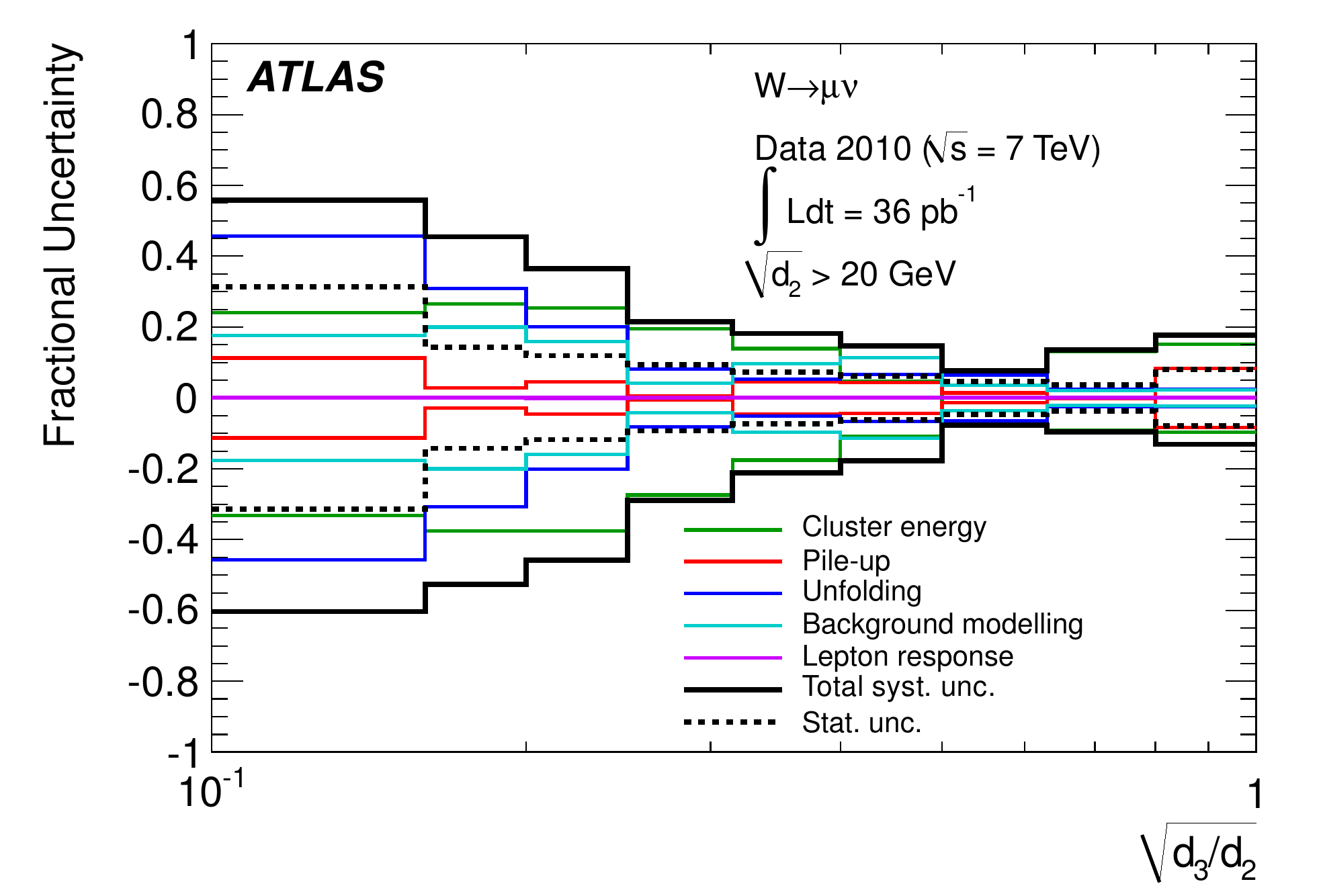}
      \caption{Summary of the systematic uncertainties on the measured particle-level ratios for $\sqrt{d_{1}/d_{0}}$ (top) and $\sqrt{d_{3}/d_{2}}$ (bottom) in the $W\to e\nu$ (left) and $W\to \mu\nu$ (right) channels.}
      \label{figure_overallunc_y13}
    \end{center}
  \end{figure*}

The magnitudes of the separate uncertainties for the hardest and fourth-hardest splittings are summarised in Figs.~\ref{figure_overallunc_d03} and \ref{figure_overallunc_y13}, where the statistical errors are also shown. Other cases are available in Appendix~\ref{sec:addsyst}. The cluster energy scale, pileup, and the unfolding procedure are the dominant sources of uncertainty in both the electron and muon channels. 

For each uncertainty an error band was calculated, whe\-re the upper limit is defined as the variation leading to larger values compared to the nominal distribution and the lower limit as the variation leading to lower values. To avoid underestimating the uncertainty in bins where statistical fluctuations were large, if both variations led to a shift in the same direction the larger difference with respect to the nominal distribution was taken as a symmetric uncertainty. Correlations between separate sources of systematic uncertainties and between different bins of the distributions were not considered. The quadratic sum of all systematic uncertainties considered above was taken to be the overall systematic uncertainty on the distributions. The overall systematic uncertainty ranges between 10$\%$ and 60$\%$ for the $\sqrt{d_{k}}$ distributions, being largest for small splitting scales and in the vicinity of $\sqrt{d_{k}} \approx 15\gev$. The uncertainty is smallest in the vicinity of $\sqrt{d_{k}} \approx 10\gev$ as this corresponds to the peak of the distribution and is thus less sensitive to scale uncertainties. For the $\sqrt{d_{k+1}/d_{k}}$ ratio distributions the overall systematic uncertainty ranges between 5$\%$ and 95$\%$, being largest for small values of the ratios. The statistical uncertainty on the unfolded measurement was combined in quadrature with the systematic uncertainty to obtain the total uncertainty.

\section{Results}
\label{section_results}
\begin{figure*}[tbp]
\begin{center}
\hfill\includegraphics[width=0.45\textwidth]{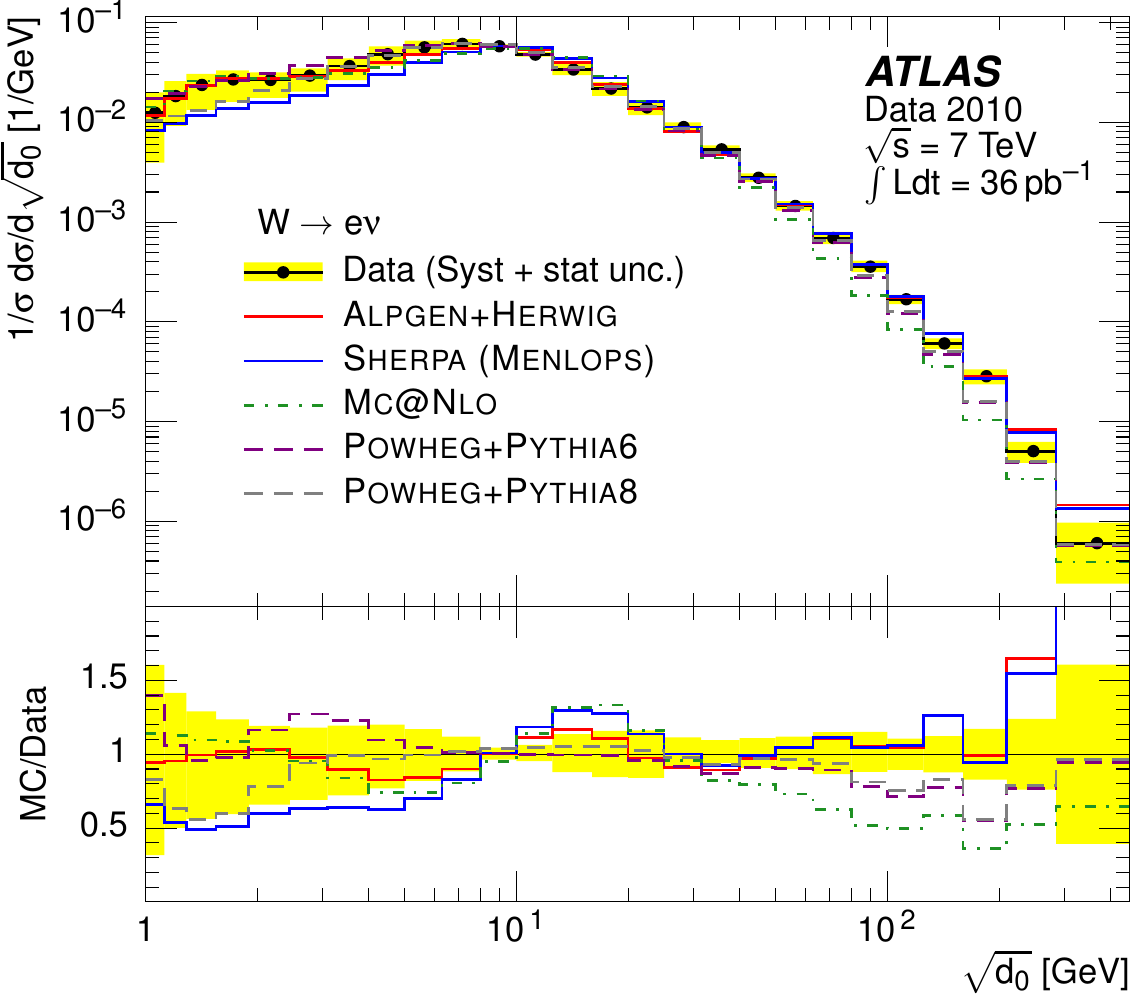}
\hfill\includegraphics[width=0.45\textwidth]{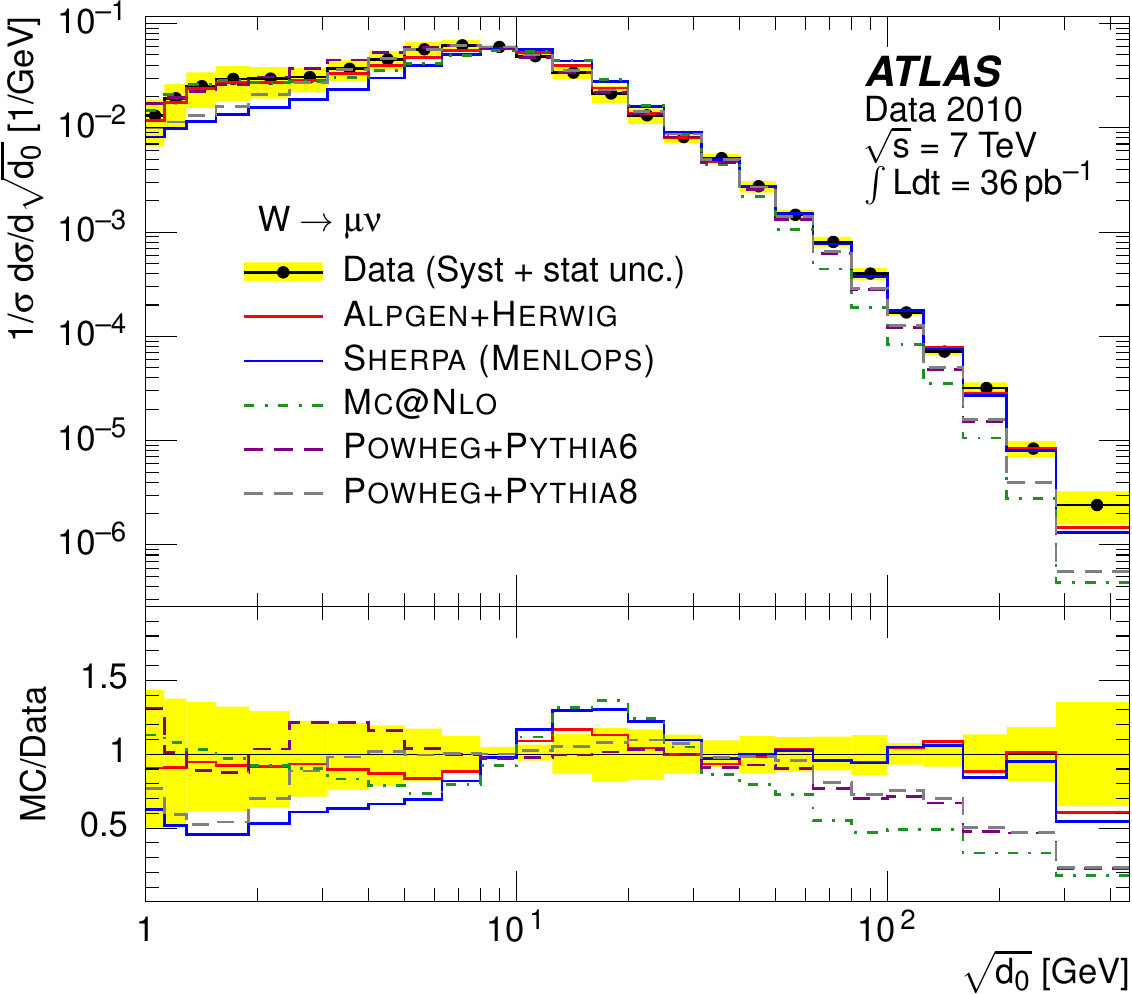}
\hfill\vphantom{.}\\\vspace*{2mm}
\hfill\includegraphics[width=0.45\textwidth]{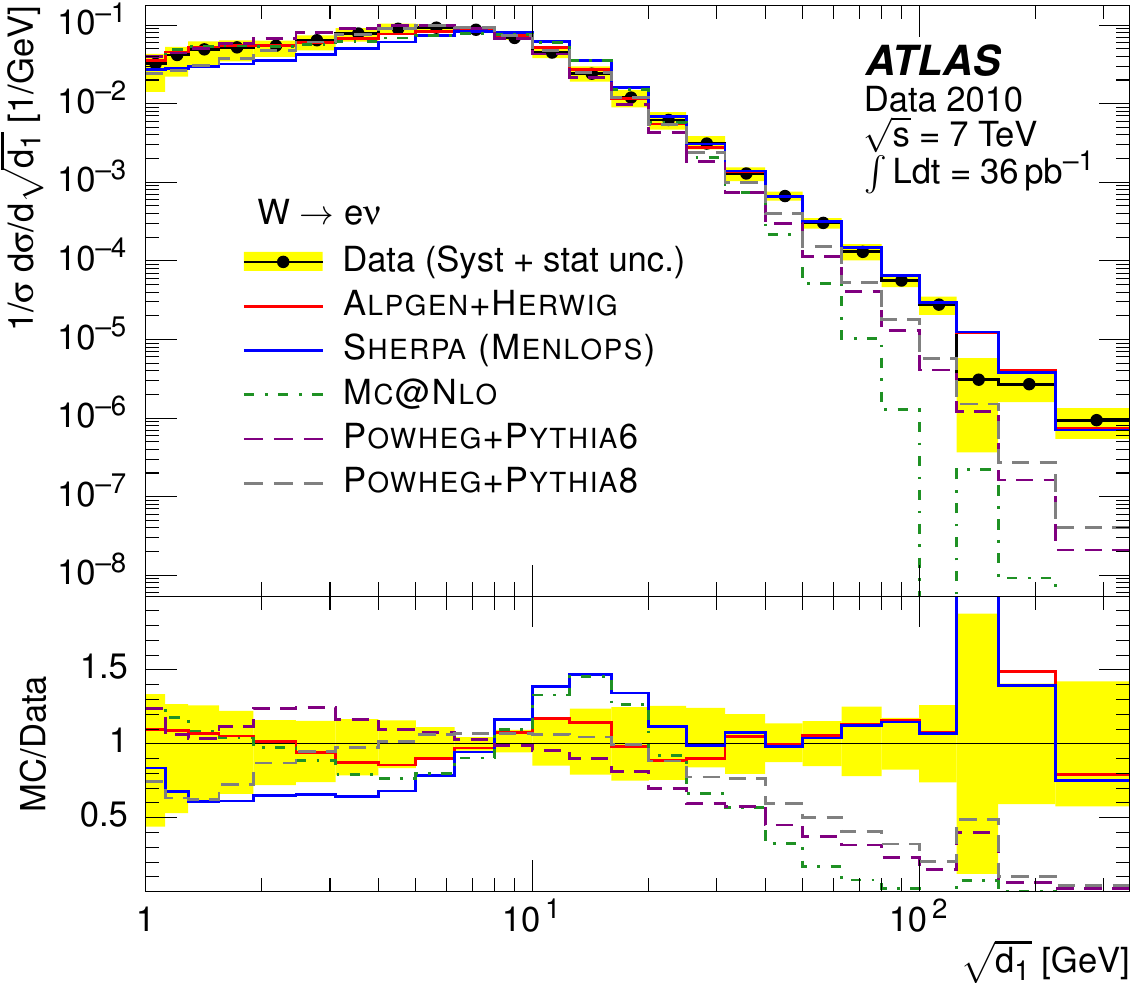}
\hfill\includegraphics[width=0.45\textwidth]{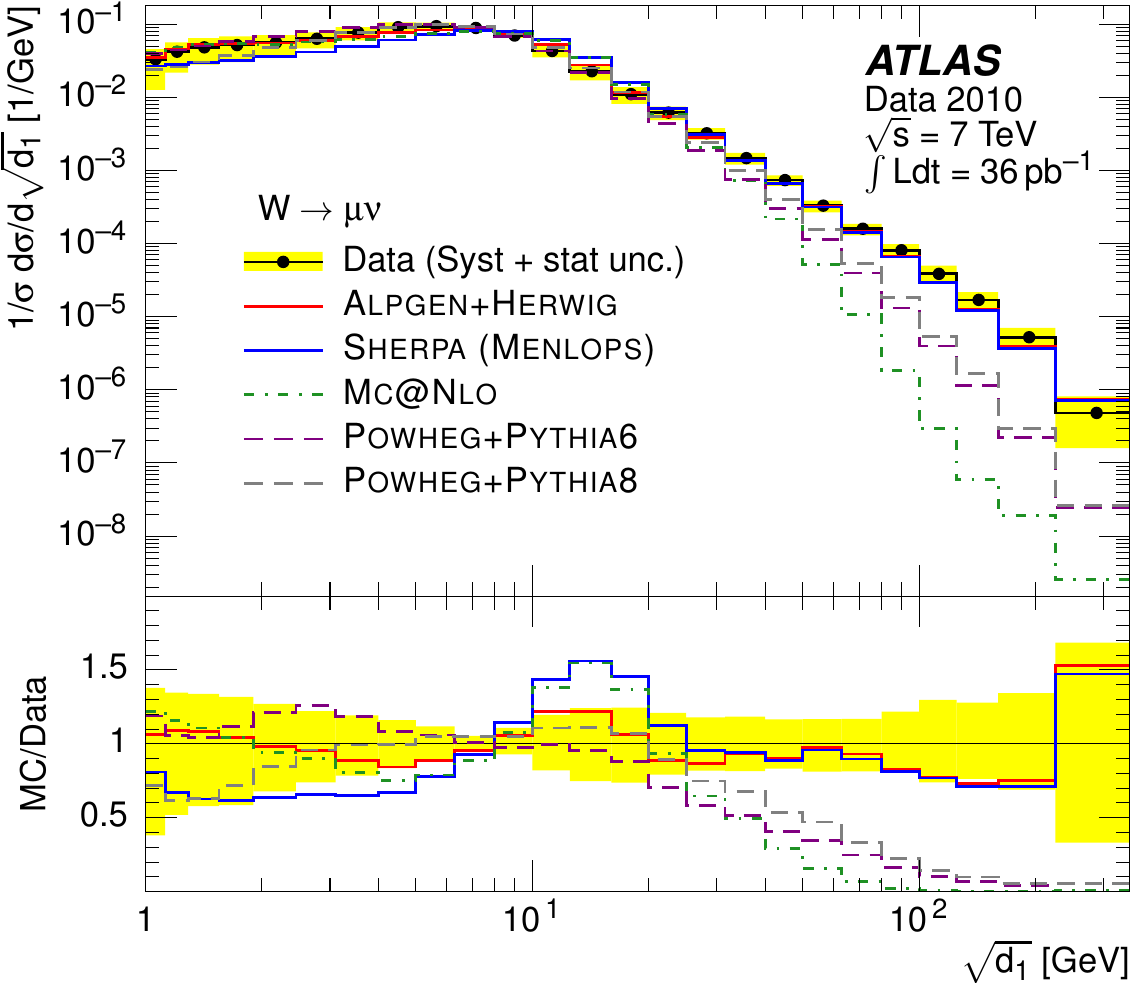}
\hfill\vphantom{.}
\caption{\label{figure_unfolded01} Distributions of $\sqrt{d_{0}}$ (top) and $\sqrt{d_{1}}$ (bottom) in the $W\to e\nu$ (left) and $W\to \mu\nu$ (right) channels, shown at particle level. The data (markers) are compared to the predictions from various MC generators, and the shaded bands represent the quadrature sum of systematic and statistical uncertainties on each bin. The histograms have been normalised to unity.}
\end{center}
\end{figure*}
\begin{figure*}[t]
\begin{center}
\hfill\includegraphics[width=0.45\textwidth]{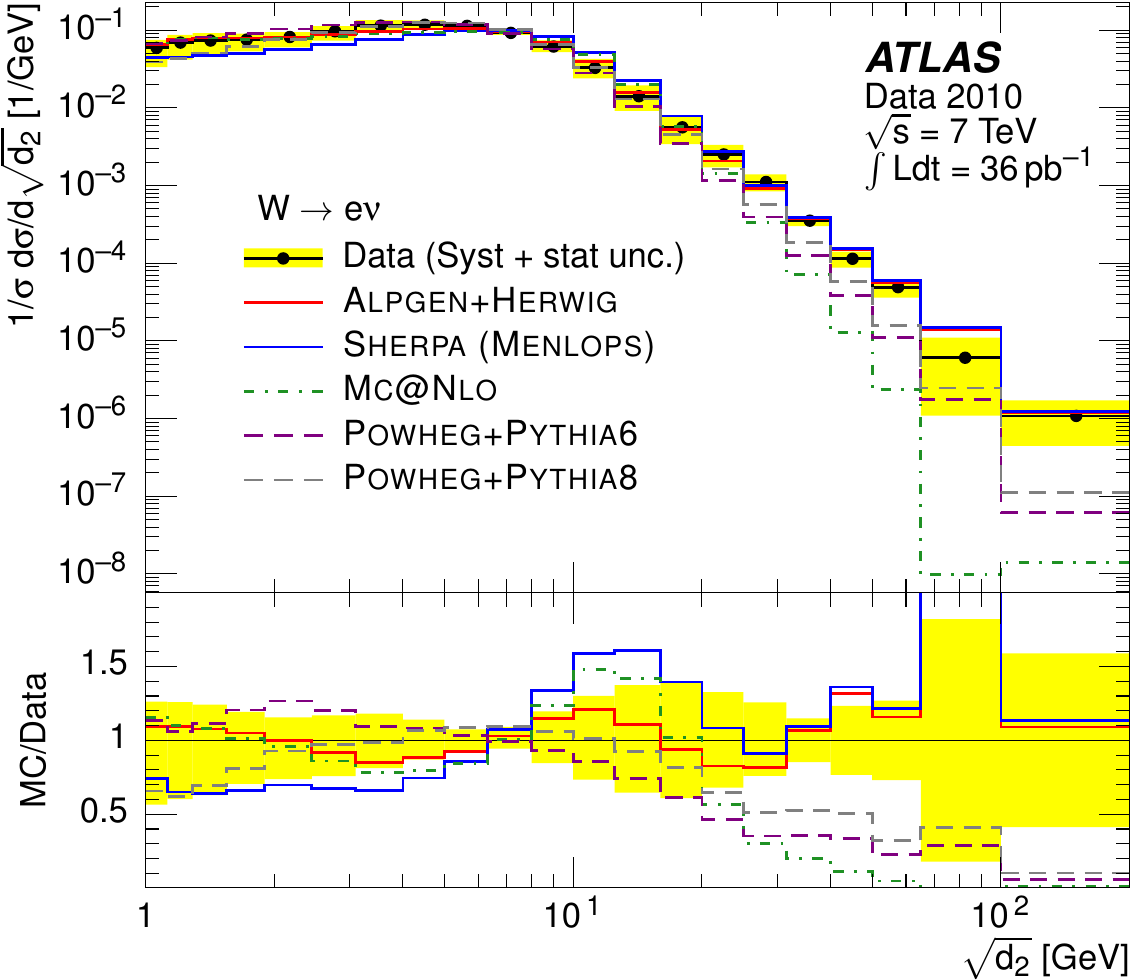}
\hfill\includegraphics[width=0.45\textwidth]{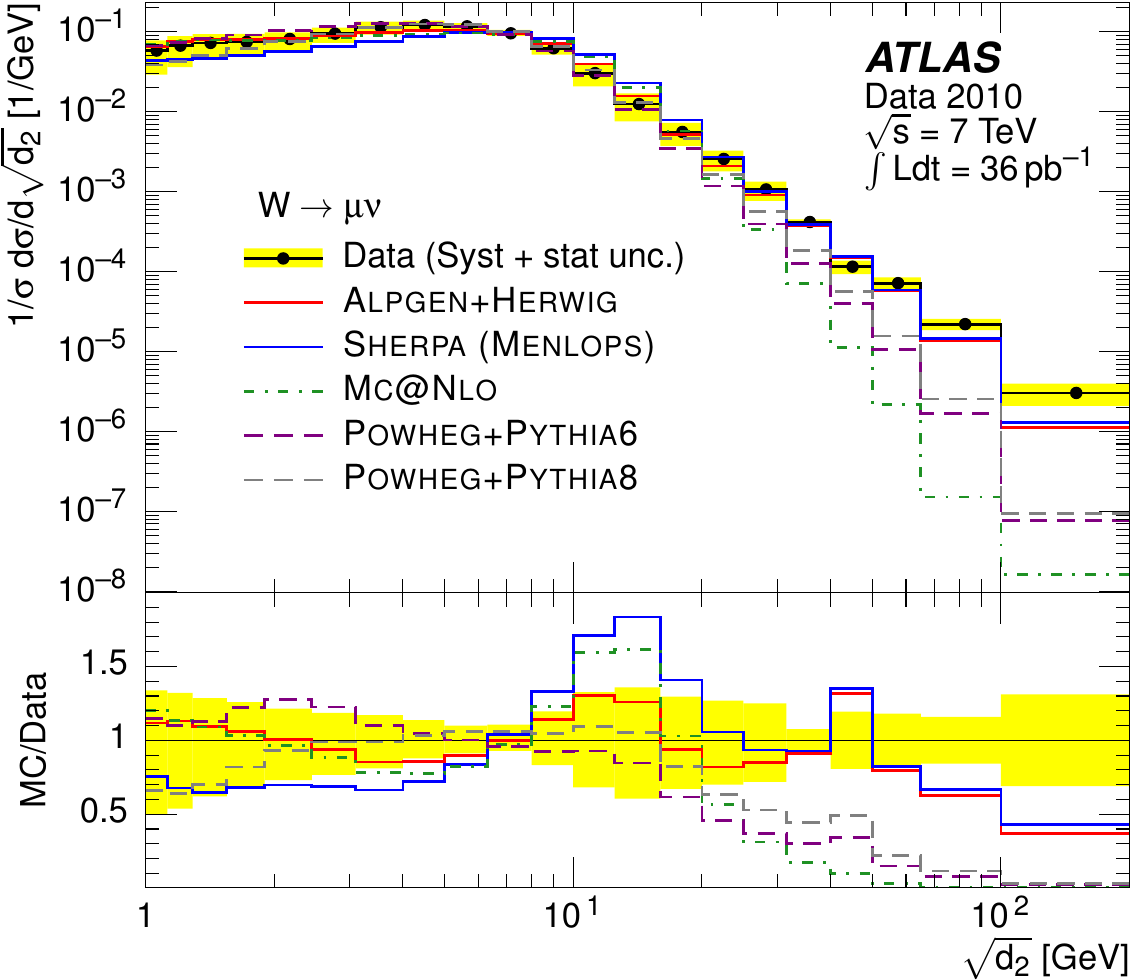}
\hfill\vphantom{.}\\\vspace*{2mm}
\hfill\includegraphics[width=0.45\textwidth]{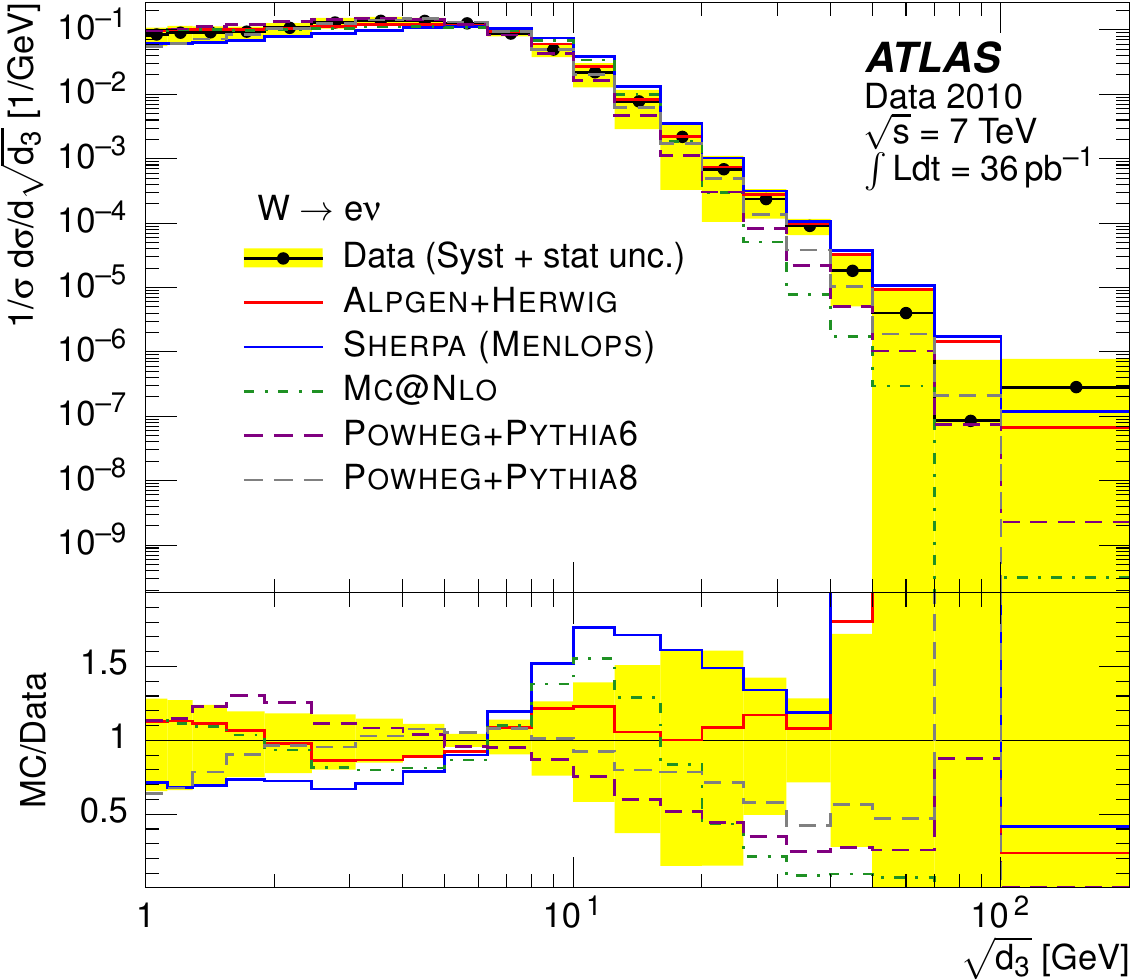}
\hfill\includegraphics[width=0.45\textwidth]{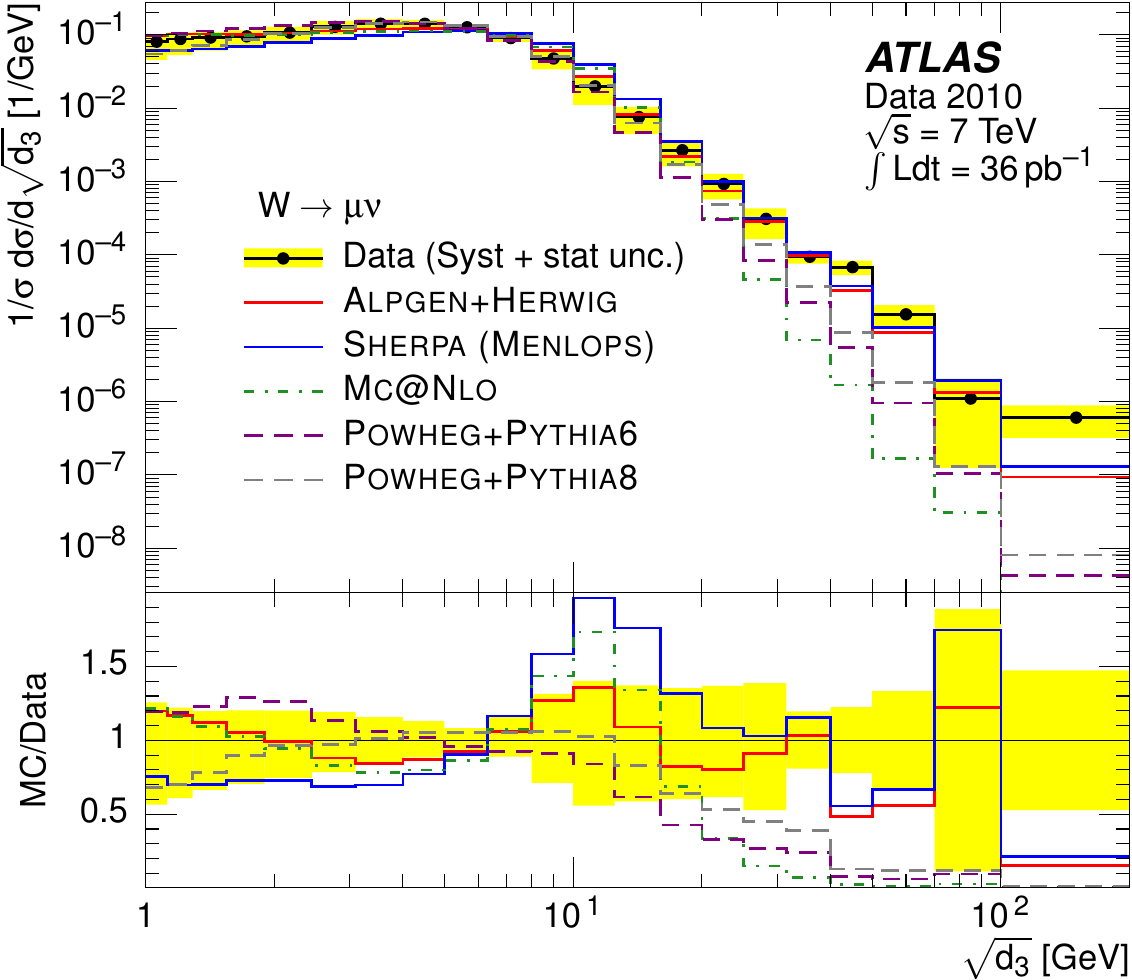}
\hfill\vphantom{.}
\caption{\label{figure_unfolded23} Distributions of $\sqrt{d_{2}}$ (top) and $\sqrt{d_{3}}$ (bottom) in the $W\to e\nu$ (left) and $W\to \mu\nu$ (right) channels, shown at particle level. The data (markers) are compared to the predictions from various MC generators, and the shaded bands represent the quadrature sum of systematic and statistical uncertainties on each bin. The histograms have been normalised to unity.}
\end{center}
\end{figure*}

\begin{figure*}[p]
\begin{center}
\hfill\includegraphics[width=0.45\textwidth]{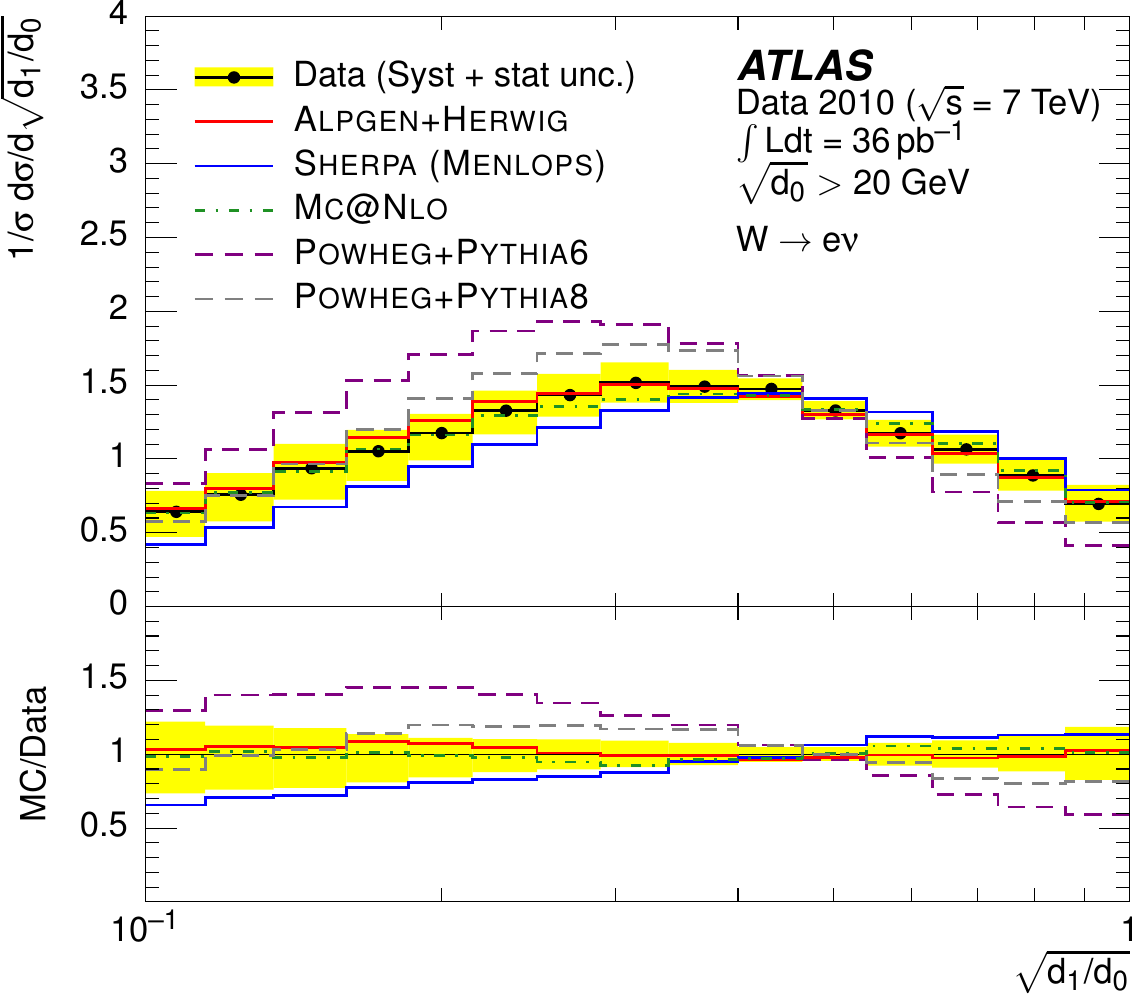}
\hfill\includegraphics[width=0.45\textwidth]{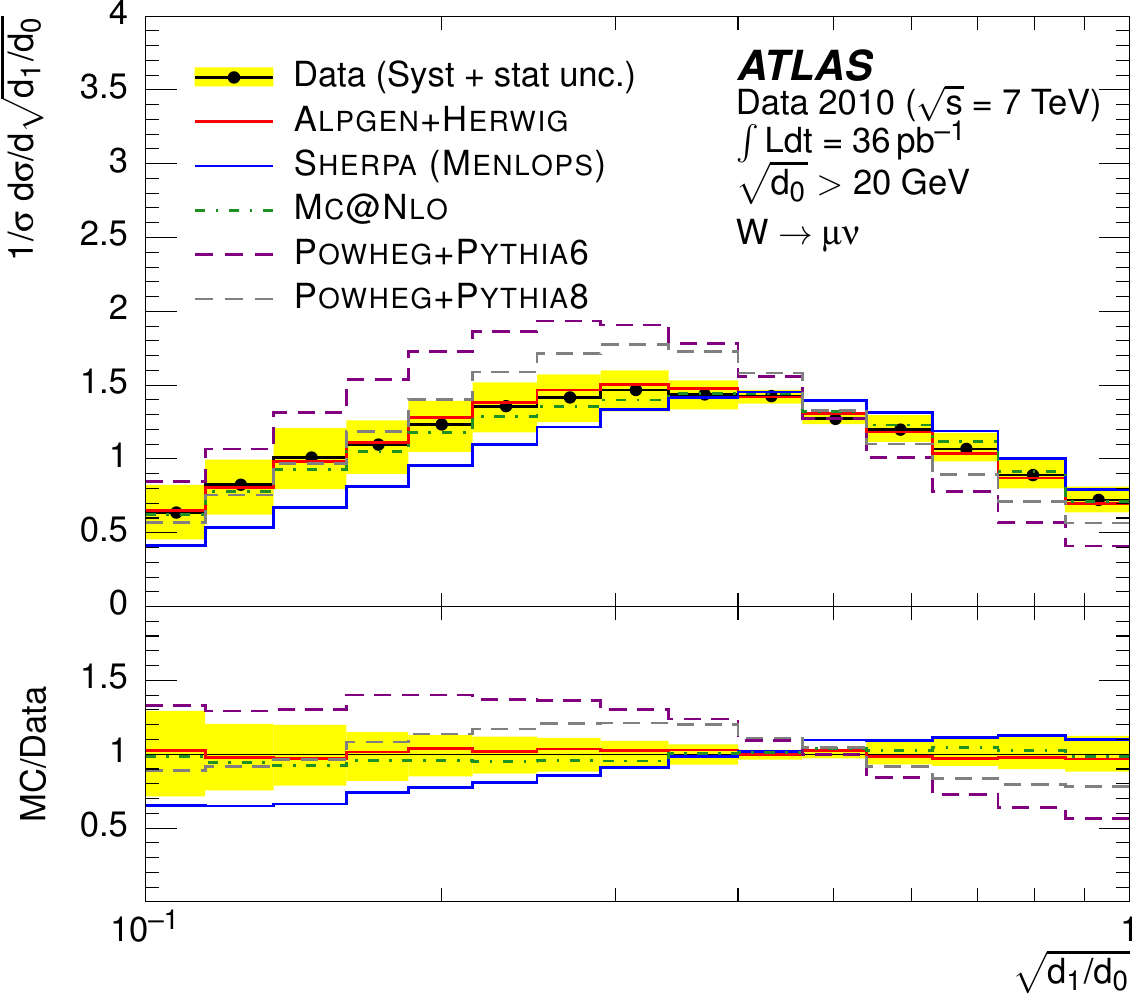}
\hfill\vphantom{.}\\\vspace*{2mm}
\hfill\includegraphics[width=0.45\textwidth]{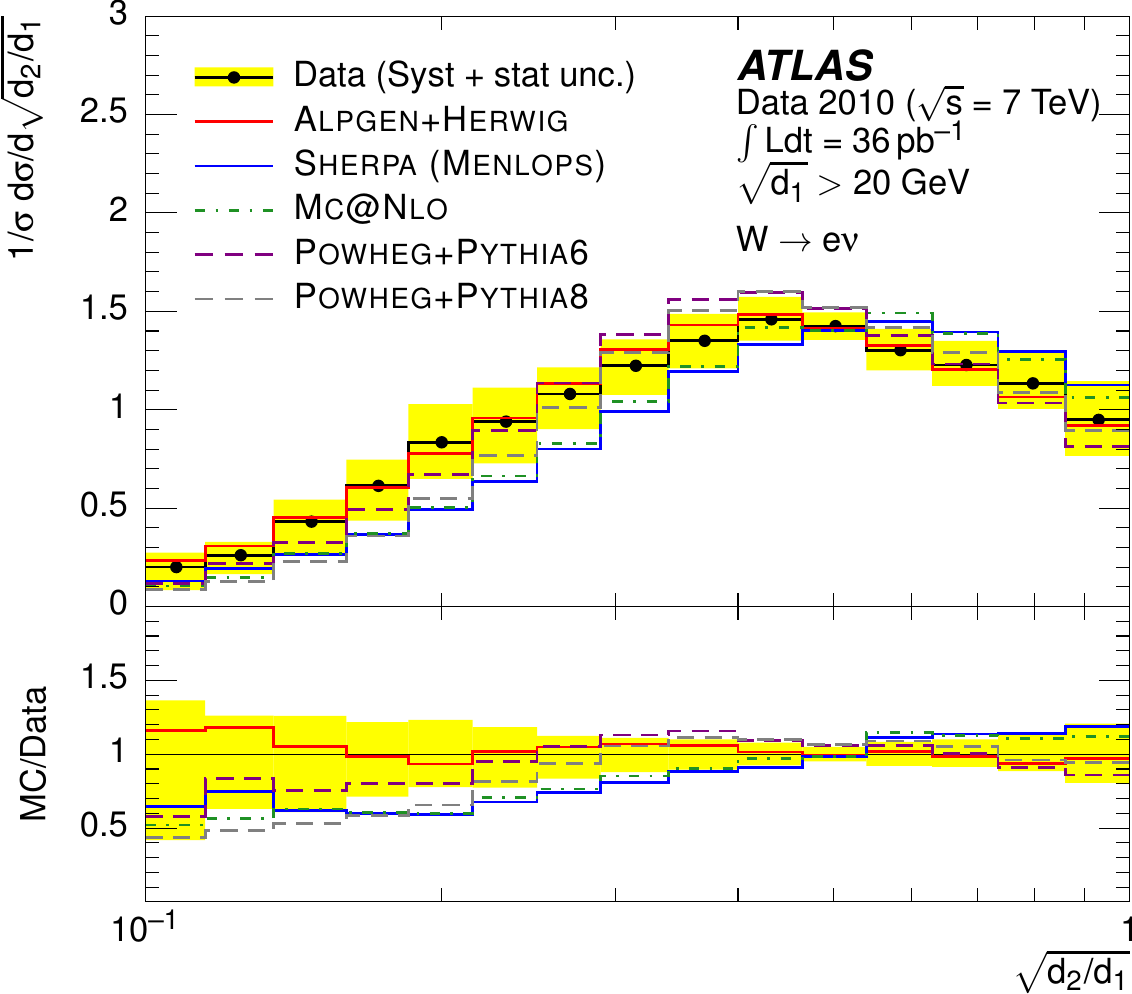}
\hfill\includegraphics[width=0.45\textwidth]{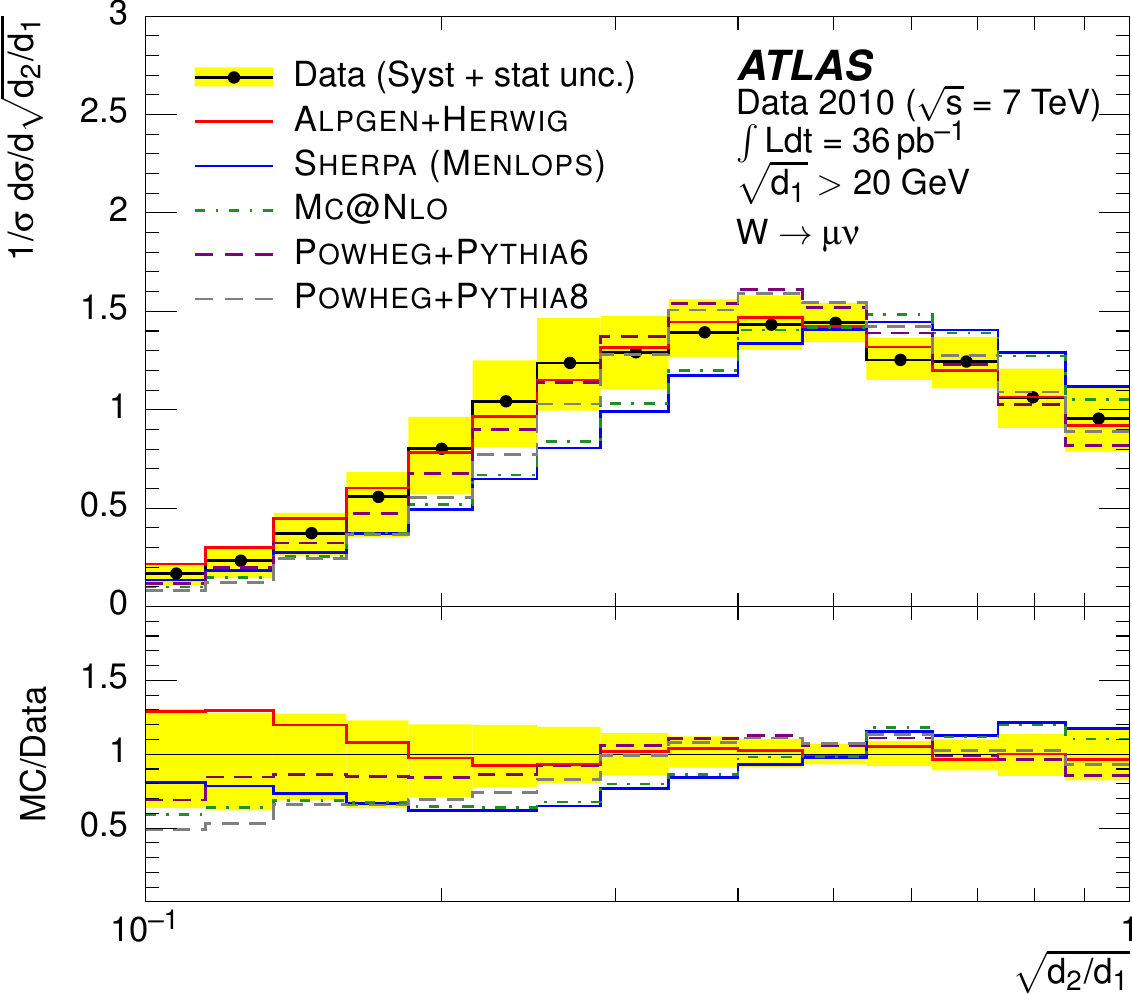}
\hfill\vphantom{.}\\\vspace*{2mm}
\hfill\includegraphics[width=0.45\textwidth]{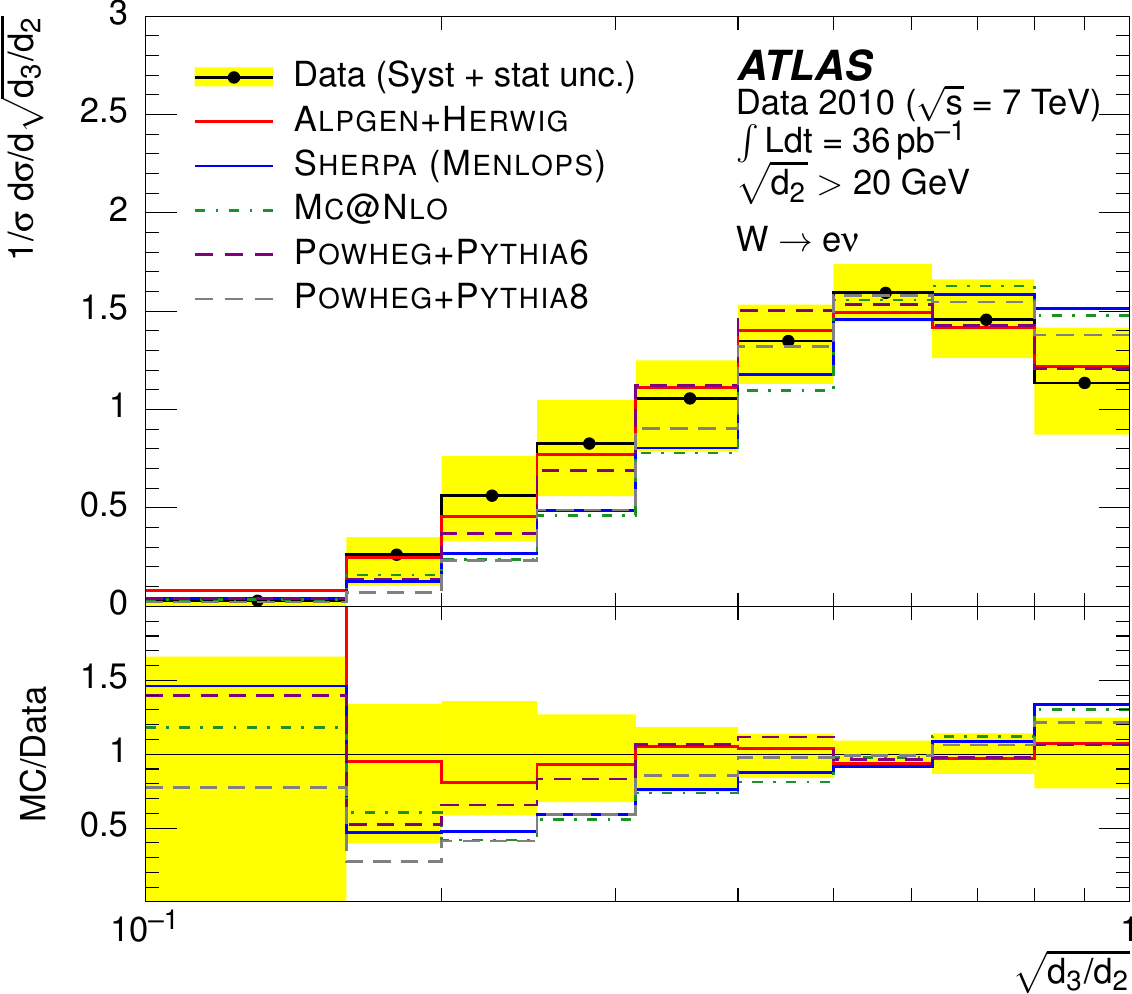}
\hfill\includegraphics[width=0.45\textwidth]{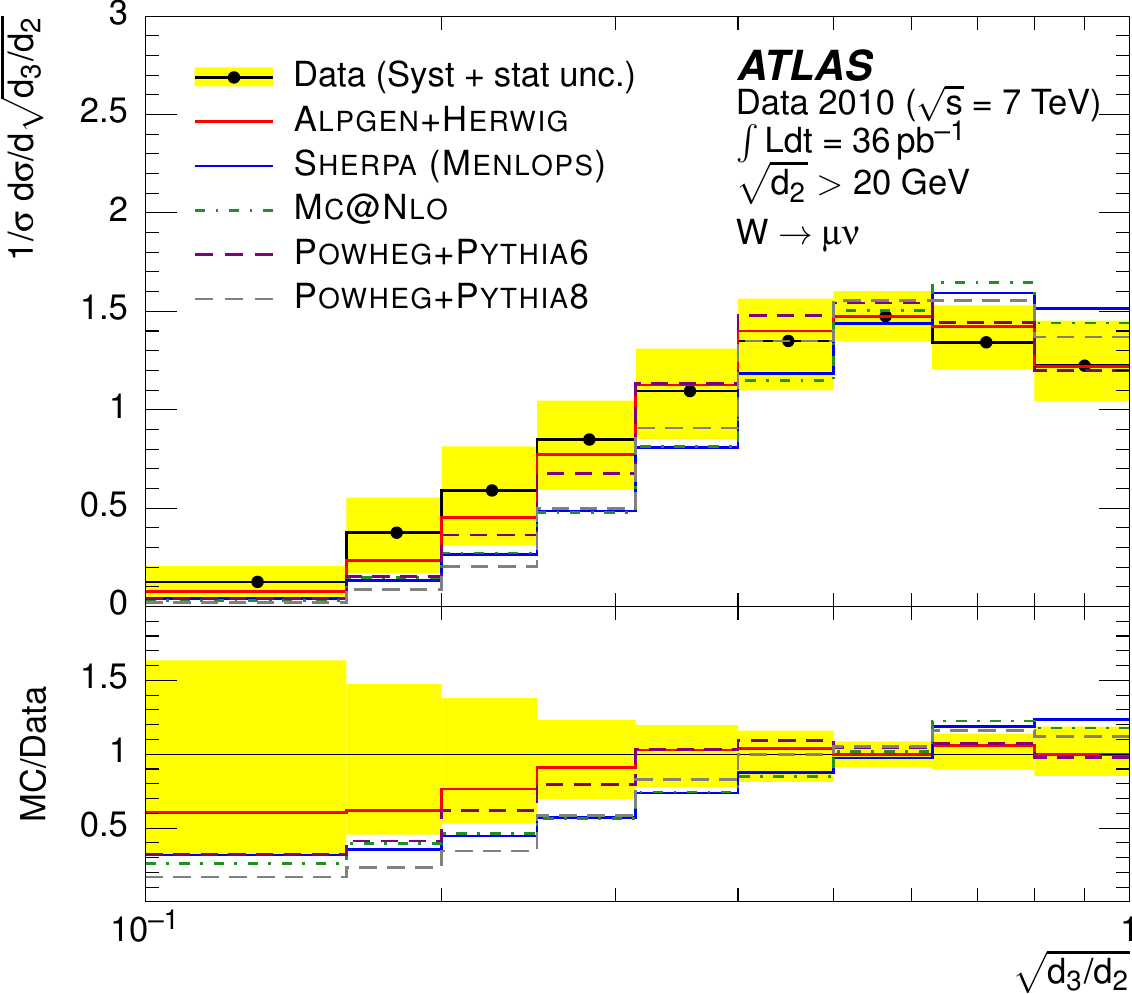}
\hfill\vphantom{.}
\caption{\label{figure_unfoldedratio}Distributions of the $\sqrt{d_{k+1}/d_{k}}$ ratio distributions for $W\to e\nu$ (left) and $W\to \mu\nu$ (right) in the data after correcting to particle level (marker) in comparison with various MC generators as described in the text. The shaded bands represent the quadrature sum of systematic and statistical uncertainties on each bin. The histograms have been normalised to unity.}
\end{center}
\end{figure*}

The different MC simulations in Sect.~\ref{sec:mc} were compared to the data using Rivet~\cite{Buckley:2010ar}.
The FastJet library~\cite{Cacciari:2011ma} was used to construct the \kt cluster sequence.
Figures~\ref{figure_unfolded01} and \ref{figure_unfolded23} display the $\sqrt{d_k}$ distributions, which have been individually norma\-lised to unity to allow for shape comparisons.

The \alpgen MC simulation generally agrees very well with the data, as already seen in the detector-level distributions. The discrepancies between the MC and data distributions are covered by the systematic and statistical uncertainties. The \sherpa predictions are almost identical to those from \alpgen in the hard region of the distributions, $\sqrt{d_k}>20$~GeV, where tree-level matrix elements are applied.

All three generators based on NLO+PS methods, i.e.\ \mcatnlo, \powhegsix and \powhegeight, predict significantly less hard activity than that found in data. As expected, this effect is strongest for higher multiplicities $k\geq 1$, where in NLO+PS generators no matrix elements are used for the description of the QCD emission. It is interesting that they also do not describe well the hard tail of the hardest splitting scale $\sqrt{d_0}$, even though they are nominally at the same leading-order accuracy as \alpgen and \sherpa in this distribution.
This may be due to differences in higher-multiplicity parton processes becoming relevant in that region or different scale choices in the real-emission matrix element or a combination of both.

In the intermediate region of $10$--$20 \gev$, both \sherpa and \mcatnlo show a similar excess over data in all $\sqrt{d_k}$. For \sherpa it is compensated by an undershoot in the very soft region, while for \mcatnlo the soft region is described well. \powhegsix and \powhegeight also agree with data in the soft region, and their deviations from each other due to the differences in parton showering and hadronisation lie within the experimental uncertainties. They give identical predictions for the hard region of $\sqrt{d_0}$, where both of them should be dominated by an identical real-emission matrix element. This confirms the expectation that the hard region is dominated by perturbative effects while resummation and non-perturbative effects have a large influence in the softer regions.

The distributions of the ratios $\sqrt{d_{k+1}/d_{k}}$ are displayed in Fig.~\ref{figure_unfoldedratio}. These probe the probability for a QCD emission of hardness $\sqrt{d_{k+1}}$ given a previous emission of scale $\sqrt{d_k}$. The \textsc{Herwig} parton shower used with both \textsc{Alpgen} and \mcatnlo gives the best description of these observables. None of the ratio observables are expected to be dominated by perturbative effects, since the bulk of the events are collected near the lower threshold at $\sqrt{d_{k}}=20$~GeV, and $\sqrt{d_{k+1}}$ is always softer than $\sqrt{d_k}$. The \textsc{Powheg} predictions, particularly for the case where \textsc{Powheg} is matched to \textsc{Pythia6}, deviate from the data in the ratio of the hardest and second-hardest clustering, $\sqrt{d_{1}/d_{0}}$.  This is the only ratio observable that directly probes the NLO+PS matching in \textsc{Powheg} and \mcatnlo.

\section{Conclusions}
\label{sec:conclusions}
A first measurement of the \kt cluster splitting scales in $W$ boson production at
a hadron--hadron collider has been presented. The measurement was performed using the 2010
data sample from $pp$ collisions at $\sqrt{s}=7$~TeV collected with the ATLAS detector at the LHC.
The data correspond to approximately 36~pb$^{-1}$ in both the electron and muon $W$-decay channels.

Results are presented for the four hardest splitting scales in a \kt cluster
sequence, and ratios of these splitting scales.
Backgrounds were subtracted and the results were corrected
for detector effects to allow a comparison to different generator predictions
at particle level.
A weighted combination was performed to optimise the precision of the measurement.
The dominant systematic uncertainties on the measurements originate from the cluster energy
scale, pileup and the unfolding procedure.

The degree of agreement between various Monte Carlo simulations with the data varies strongly
for different regions of the observables.
The hard tails of the distributions are significantly better described
by the multi-leg generators \alpgen and \sherpa, which include exact tree-level
matrix elements, than by the NLO+PS generators \mcatnlo and \textsc{Powheg}. This also holds
true for the hardest clustering, $\sqrt{d_0}$, even though it is formally predicted at the
same QCD leading-order accuracy by all of these generators.

In the soft regions of the splitting scales, larger variations
between all generators become evident. The generators based on the \textsc{Herwig} parton
shower provide a good description of the data, while the \sherpa and \textsc{Powheg+} \textsc{Pythia}
predictions do not reproduce the soft regions of the measurement well.

With this discriminating power the data thus test the
resummation shape generated by parton showers and the extent to which
the shower accuracy is preserved by the different merging and matching methods
used in these Monte Carlo simulations.


\section{Acknowledgements}

We thank CERN for the very successful operation of the LHC, as well as the
support staff from our institutions without whom ATLAS could not be
operated efficiently.
We acknowledge the support of ANPCyT, Argentina; YerPhI, Armenia; ARC,
Australia; BMWF and FWF, Austria; ANAS, Azerbaijan; SSTC, Belarus; CNPq and FAP\-ESP,
Brazil; NSERC, NRC and CFI, Canada; CERN; CONICYT, Chile; CAS, MOST and NSFC,
China; COLCIENCIAS, Colombia; MSMT CR, MPO CR and VSC CR, Czech Republic;
DNRF, DNSRC and Lundbeck Foundation, Denmark; EPLANET, ERC and NSRF, European Union;
IN2P3-CNRS, CEA-DSM/IRFU, France; GNSF, Georgia; BMBF, DFG, HGF, MPG and AvH
Foundation, Germany; GSRT and NSRF, Greece; ISF, MINERVA, GIF, DIP and Benoziyo Center,
Israel; INFN, Italy; MEXT and JSPS, Japan; CNRST, Morocco; FOM and NWO,
Netherlands; BRF and RCN, Norway; MNiSW, Poland; GRICES and FCT, Portugal; MERYS
(MECTS), Romania; MES of Russia and ROSATOM, Russian Federation; JINR; MSTD,
Serbia; MSSR, Slovakia; ARRS and MVZT, Slovenia; DST/NRF, South Africa;
MICINN, Spain; SRC and Wallenberg Foundation, Sweden; SER, SNSF and Cantons of
Bern and Geneva, Switzerland; NSC, Taiwan; TAEK, Turkey; STFC, the Royal
Society and Leverhulme Trust, United Kingdom; DOE and NSF, United States of
America.
The crucial computing support from all WLCG partners is acknowledged
gratefully, in particular from CERN and the ATLAS Tier-1 facilities at
TRIUMF (Canada), NDGF (Denmark, Norway, Sweden), CC-IN2P3 (France),
KIT/GridKA (Germany), INFN-CNAF (Italy), NL-T1 \linebreak(Netherlands), PIC (Spain),
ASGC (Taiwan), RAL (UK) and BNL (USA) and in the Tier-2 facilities
worldwide.

\clearpage
\bibliographystyle{atlasBibStyleWoTitle}
\bibliography{paper}

\clearpage

\appendix
\onecolumn

\section{Appendices}

\subsection{Additional detector-level comparisons}
\label{sec:adddists}
\begin{figure}[htbp]
\begin{center}
\includegraphics[width=0.33\textwidth]{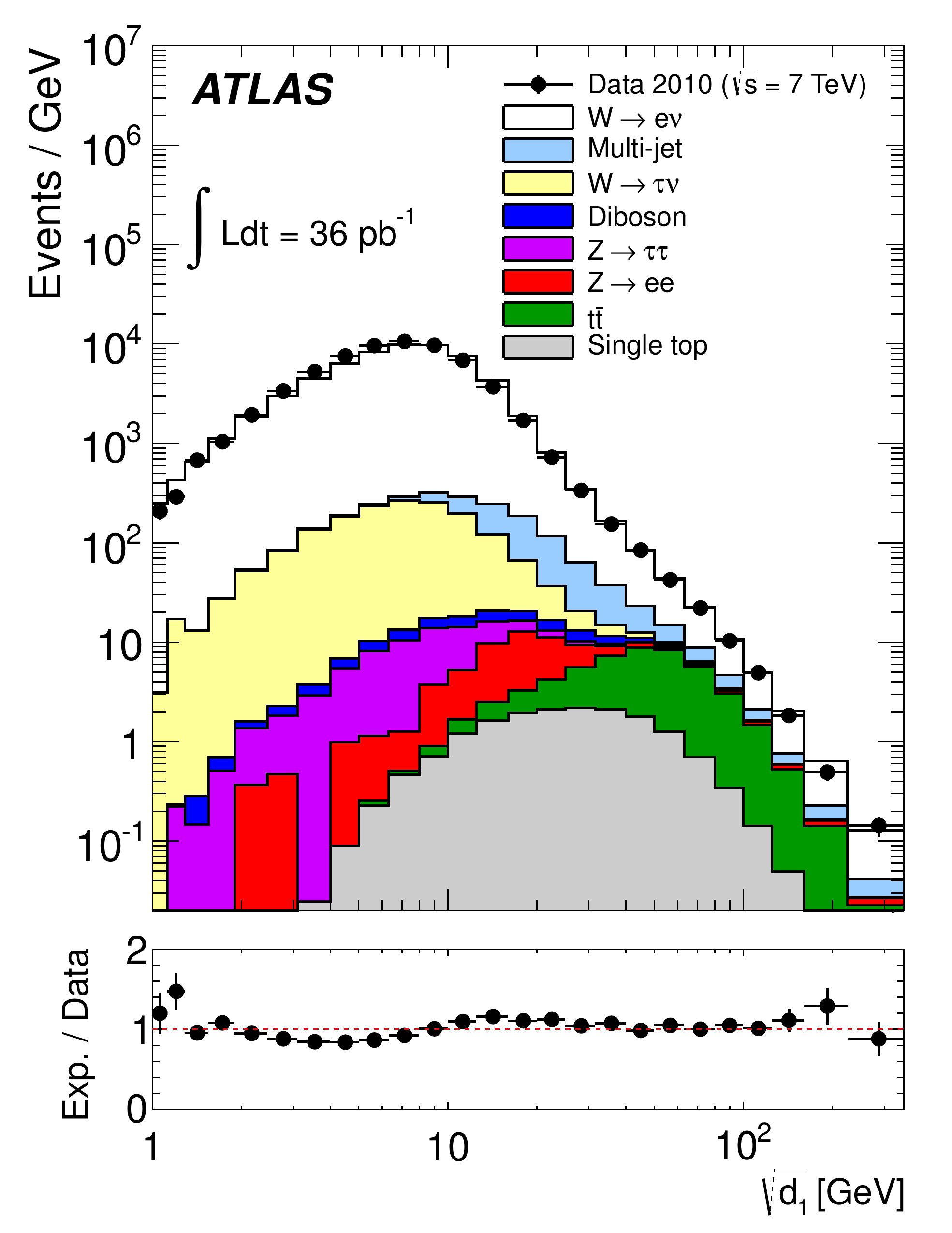}\nolinebreak
\includegraphics[width=0.33\textwidth]{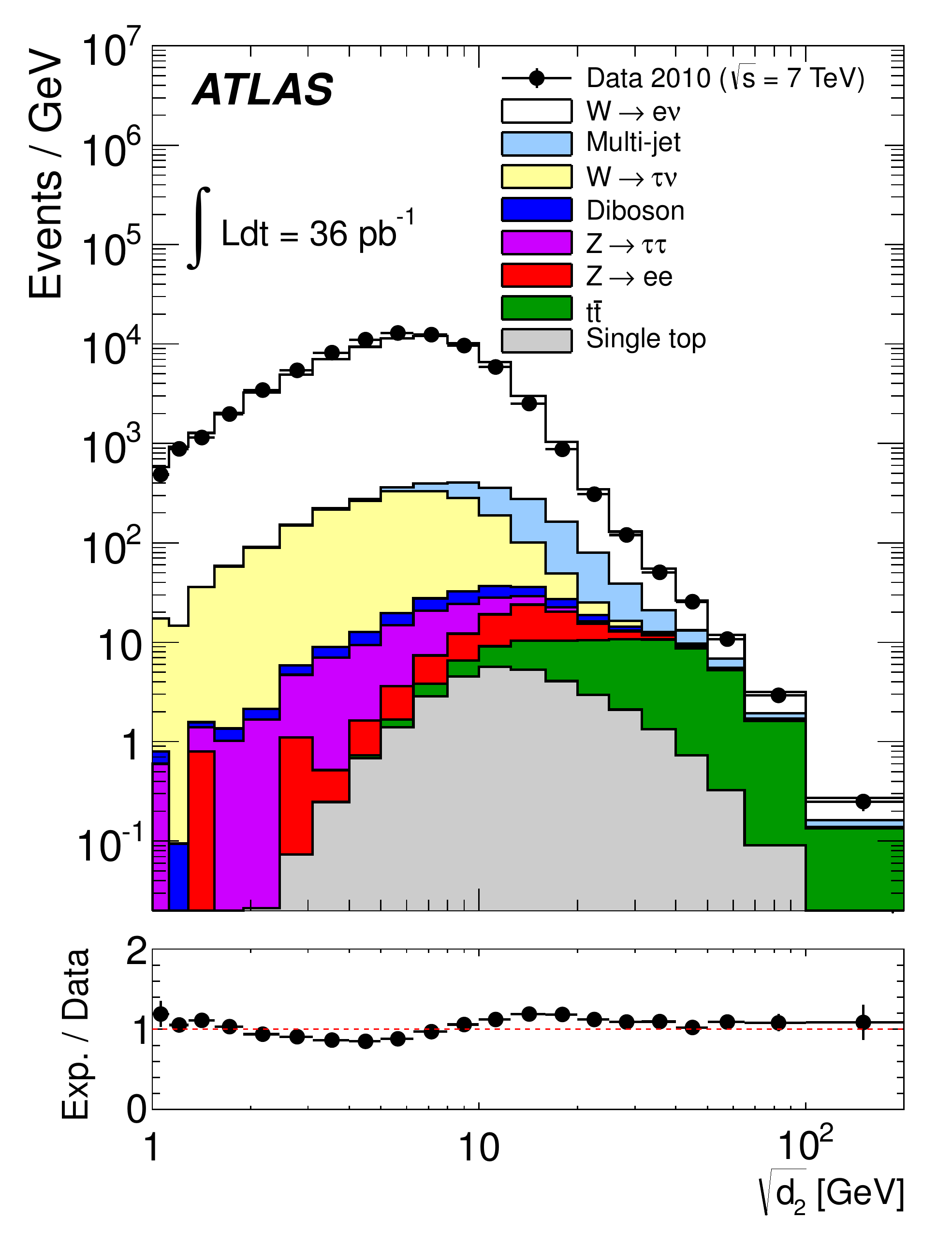}\nolinebreak
\includegraphics[width=0.33\textwidth]{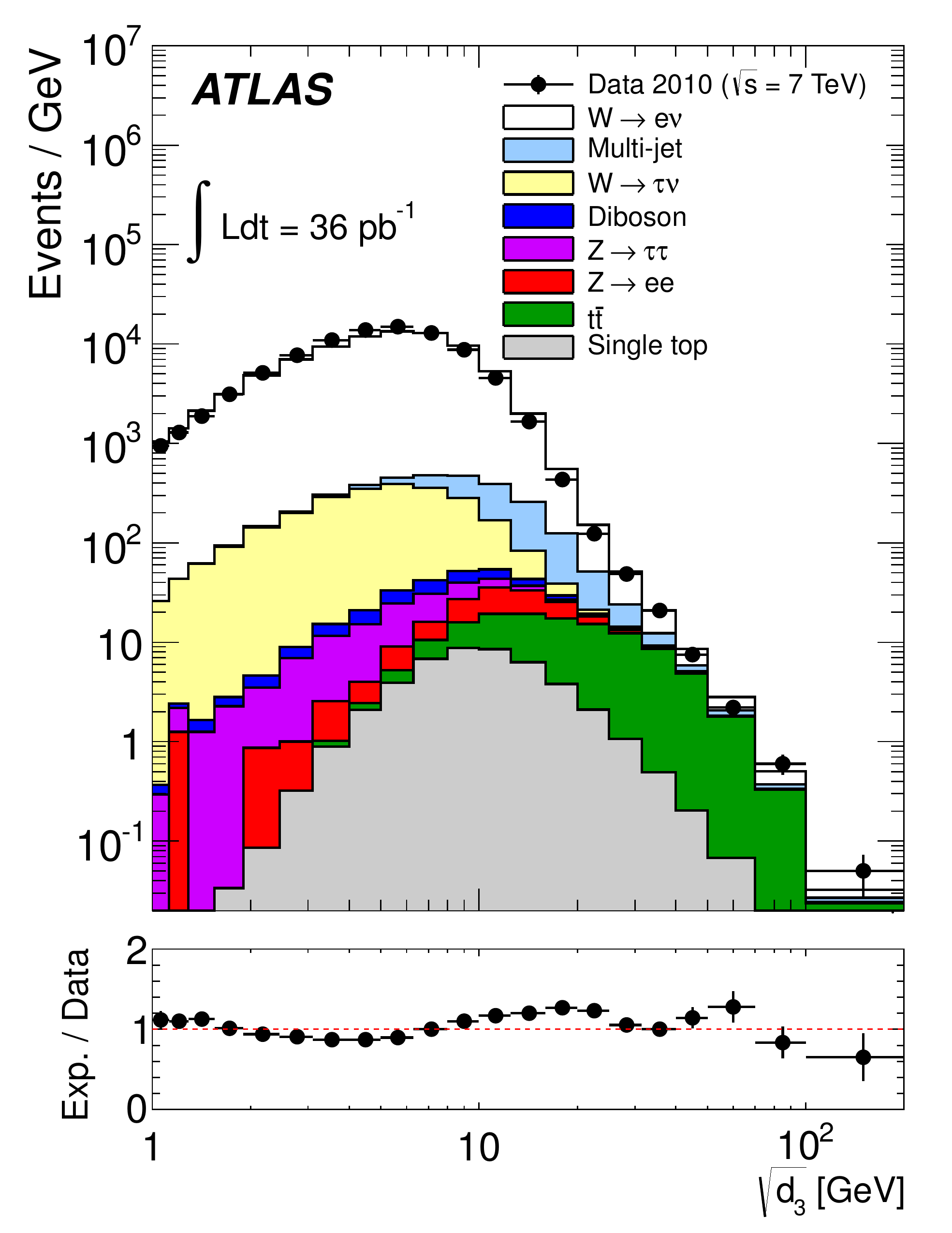}\\
\includegraphics[width=0.33\textwidth]{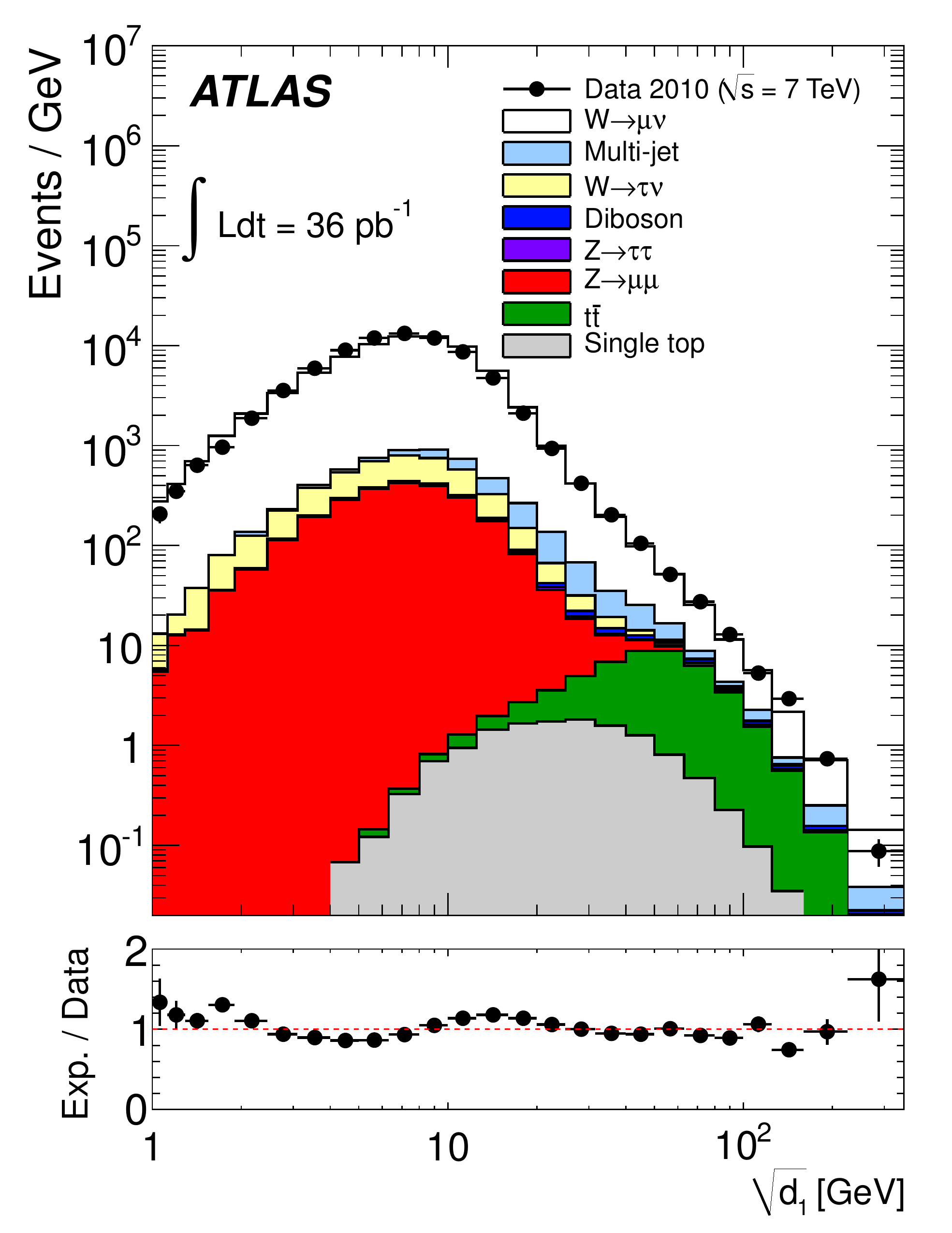}\nolinebreak
\includegraphics[width=0.33\textwidth]{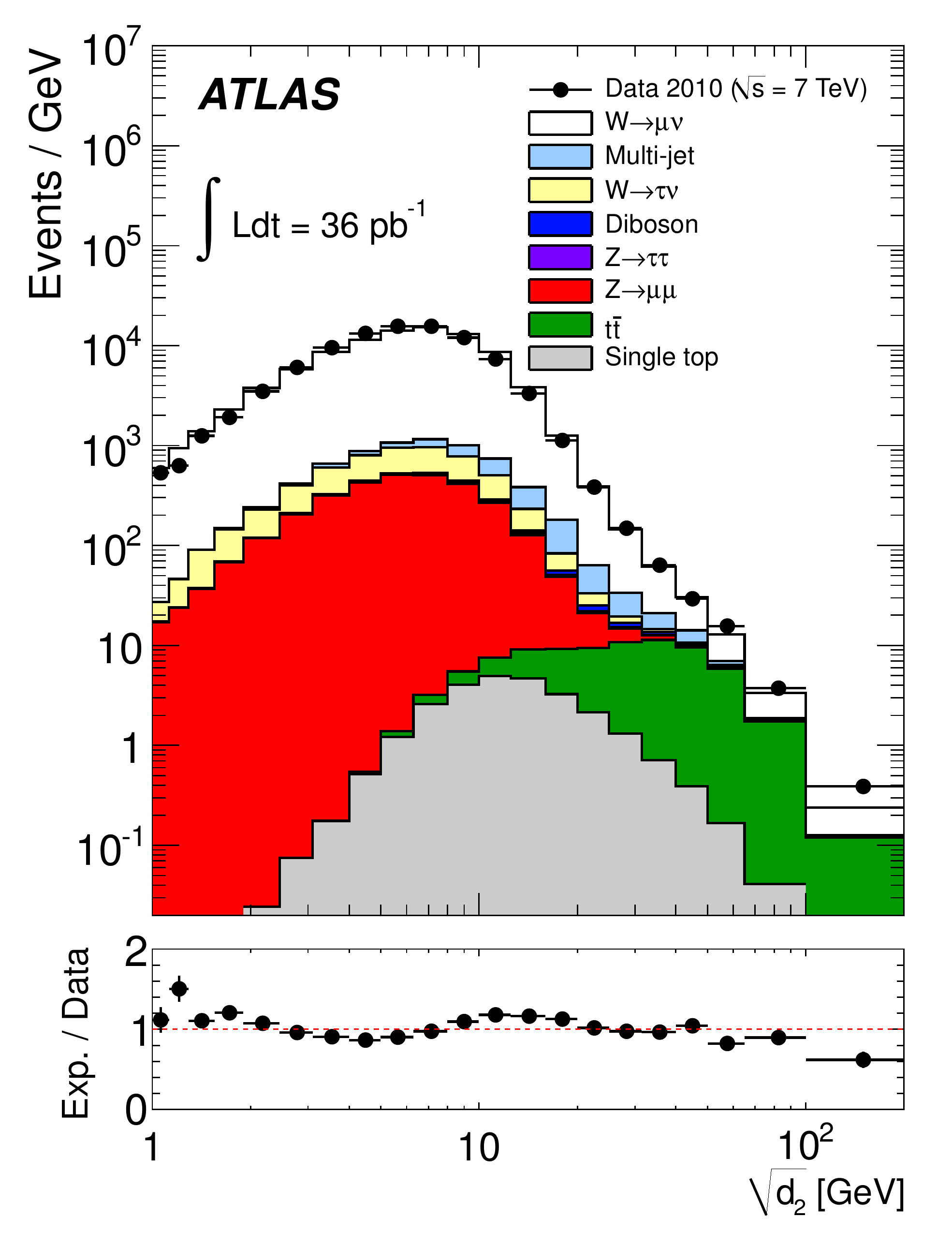}\nolinebreak
\includegraphics[width=0.33\textwidth]{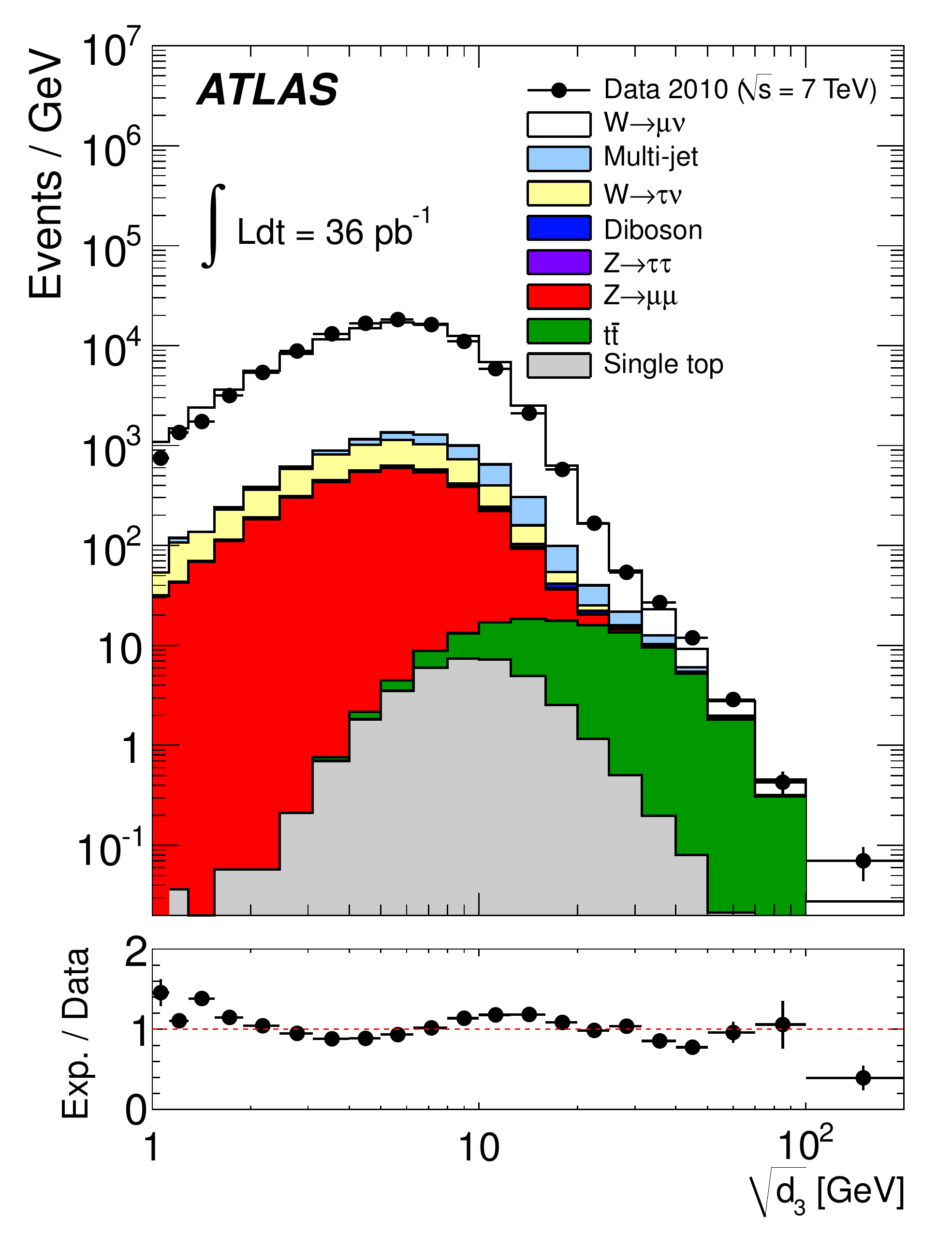}
\caption{Uncorrected splitting scales $\sqrt{d_{1}}$ (left),  $\sqrt{d_{2}}$ (middle) and $\sqrt{d_{3}}$ (right) for events passing the $W\to e\nu$ (top) and $W\to \mu\nu$ (bottom) selection requirements. The distributions from the data (markers) are compared with the predicted signal from the MC simulation, provided by \alpgen and normalised to the NNLO prediction. In addition, physics backgrounds, also shown, have been added in proportion to the predictions from the MC simulation. The ratio between the expectation and the data is shown in the lower plot. The error bars shown on the data are statistical only.}
\label{fig:figure_d123}
\end{center}
\end{figure}

\begin{figure}[htbp]  
\begin{center}
\includegraphics[width=0.4\textwidth]{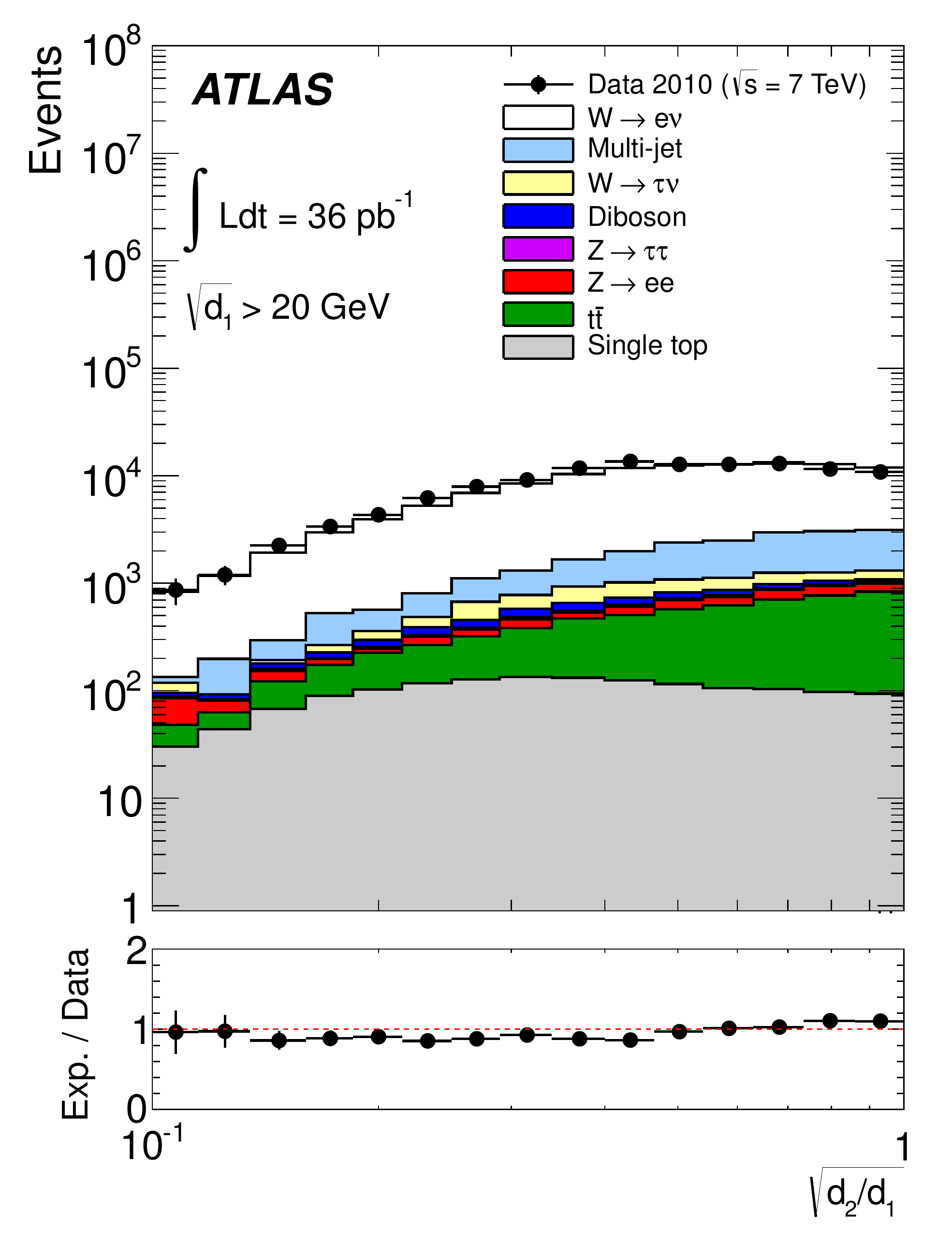}\nolinebreak
\includegraphics[width=0.4\textwidth]{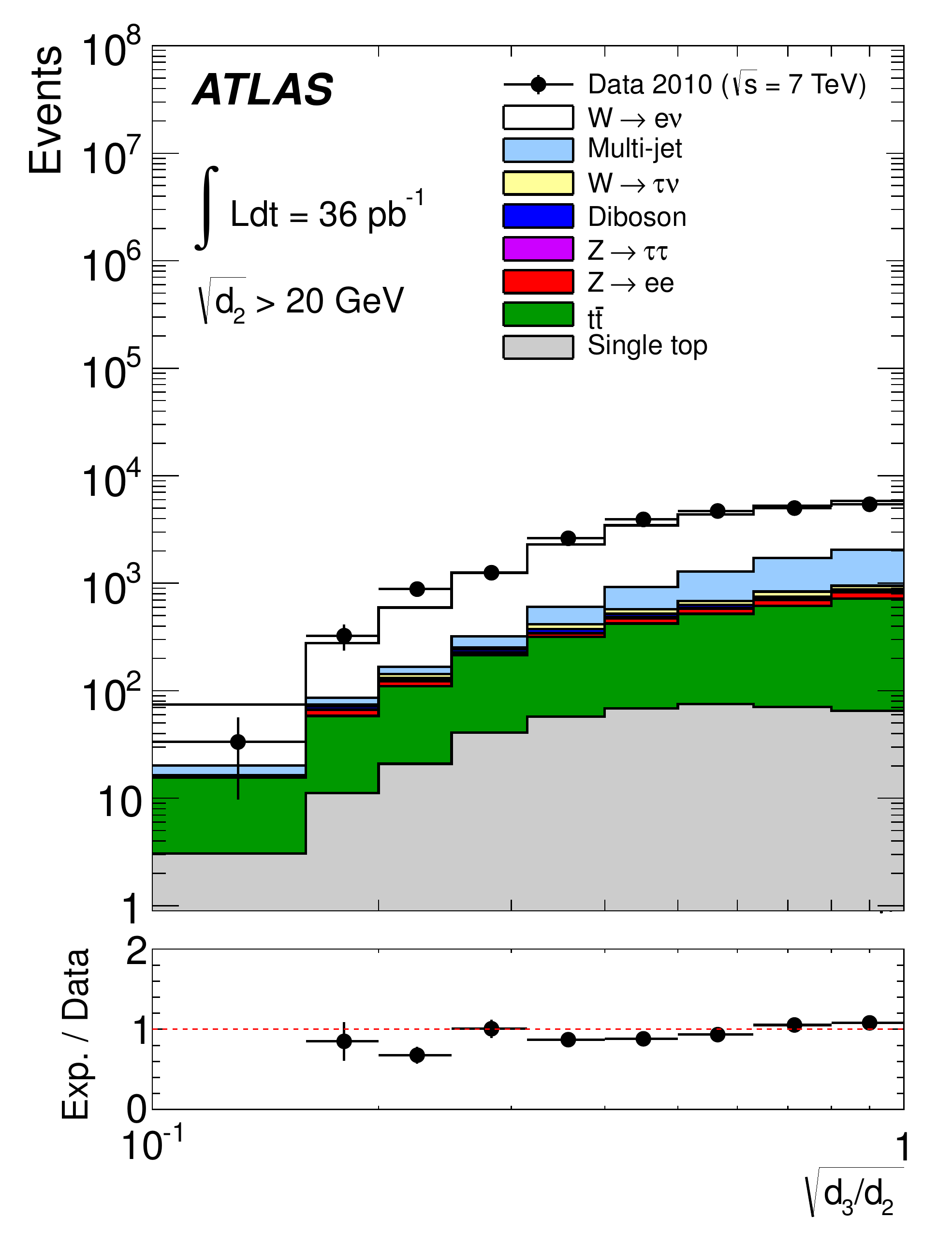}\\
\includegraphics[width=0.4\textwidth]{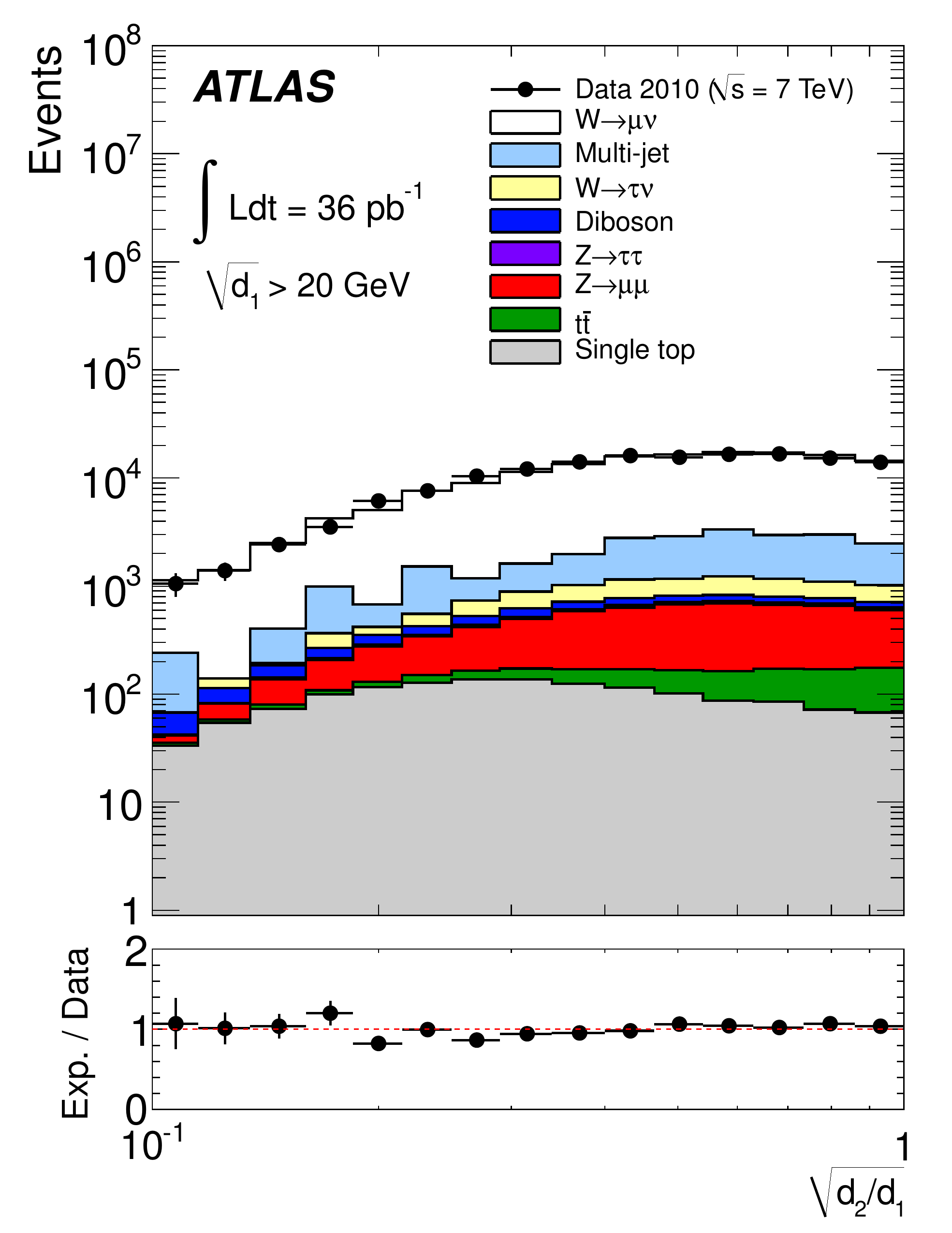}\nolinebreak
\includegraphics[width=0.4\textwidth]{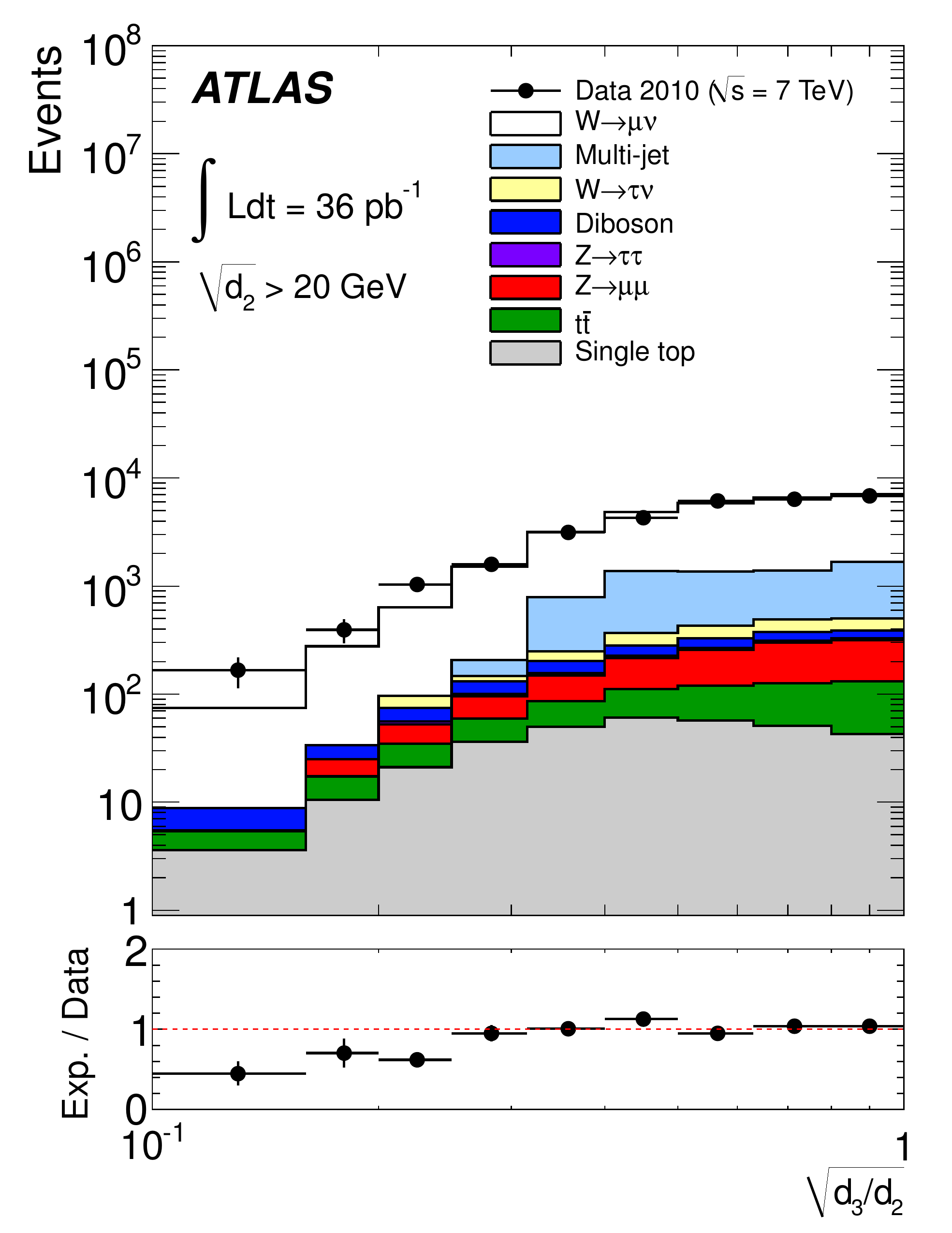}
\caption{Uncorrected ratios $\sqrt{d_{2}/d_{1}}$ (left) and $\sqrt{d_{3}/d_{2}}$ (right) for events passing the $W\to e\nu$ (top) and $W\to \mu\nu$ (bottom) selection requirements.  The distributions from the data (markers) are compared with the predicted signal from the MC simulation, provided by \alpgen and normalised to the NNLO prediction. In addition, physics backgrounds, also shown, have been added in proportion to the predictions from the MC simulation. The ratio between the expectation and the data is shown in the lower plot. The error bars shown on the data are statistical only.}
\label{fig:figure_dy23}
\end{center}
\end{figure}

\clearpage

\subsection{Additional summaries of systematic uncertainties}
\label{sec:addsyst}
\begin{figure}[h]  
\begin{center}
\includegraphics[width=0.5\textwidth]{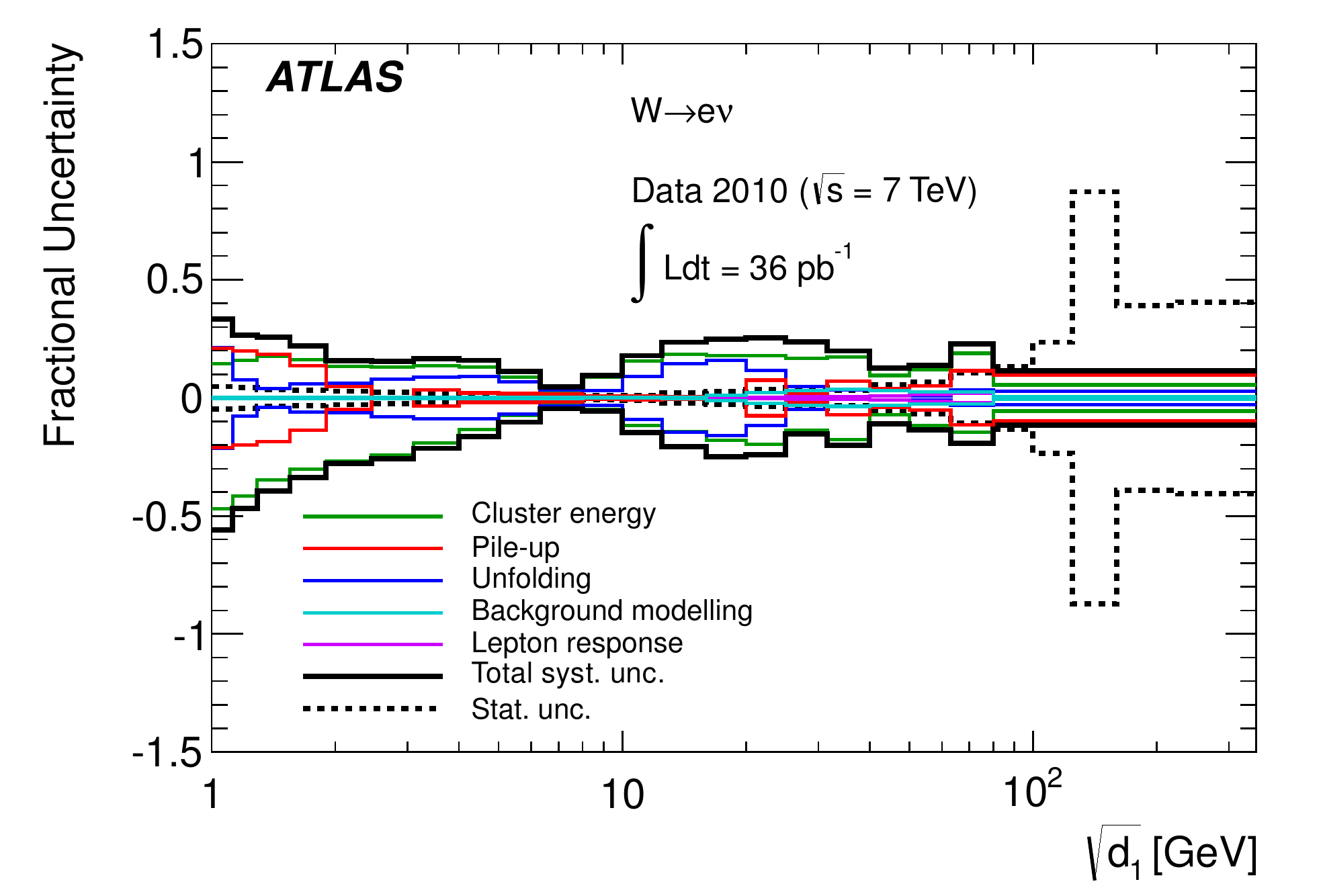}\nolinebreak
\includegraphics[width=0.5\textwidth]{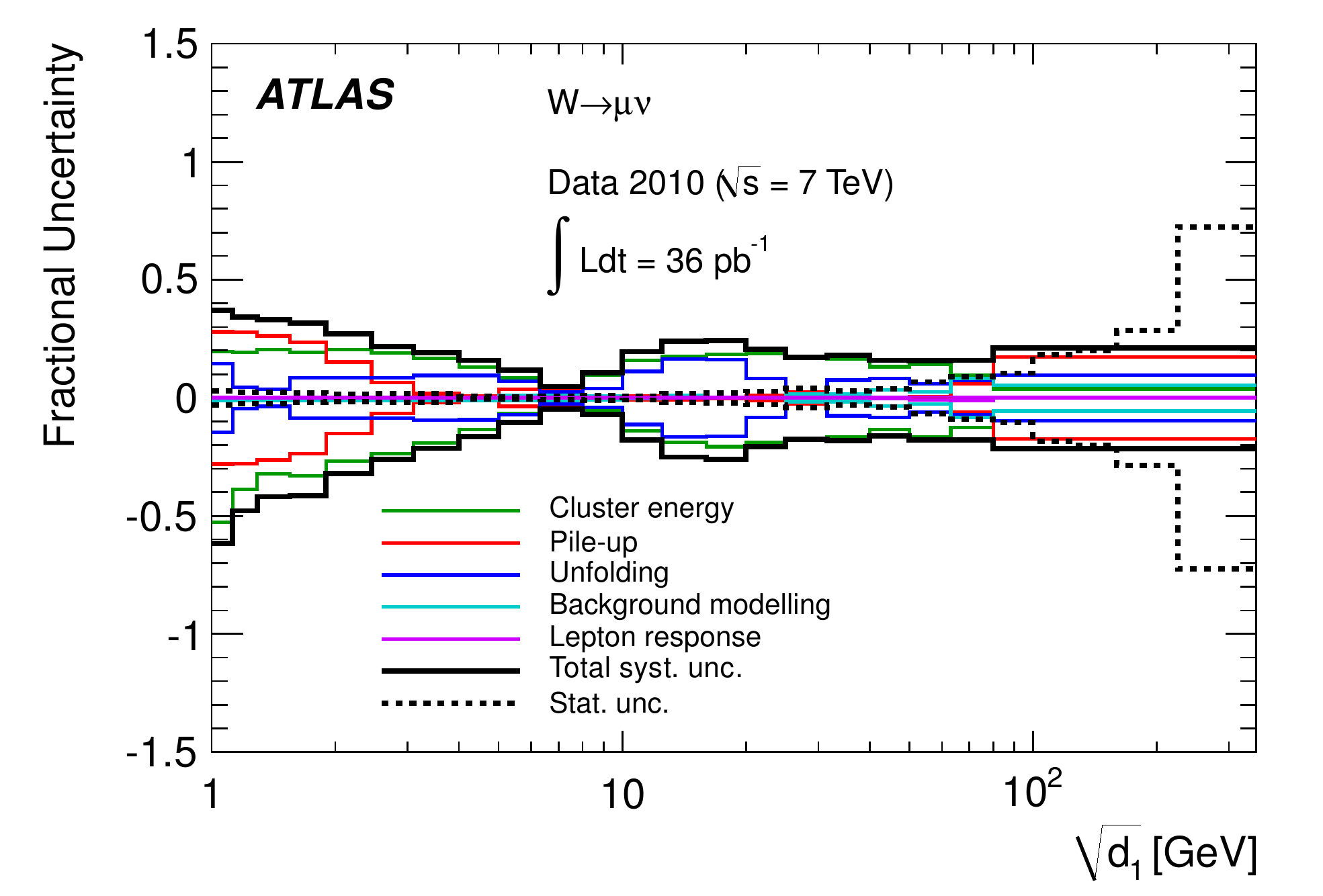}\\
\includegraphics[width=0.5\textwidth]{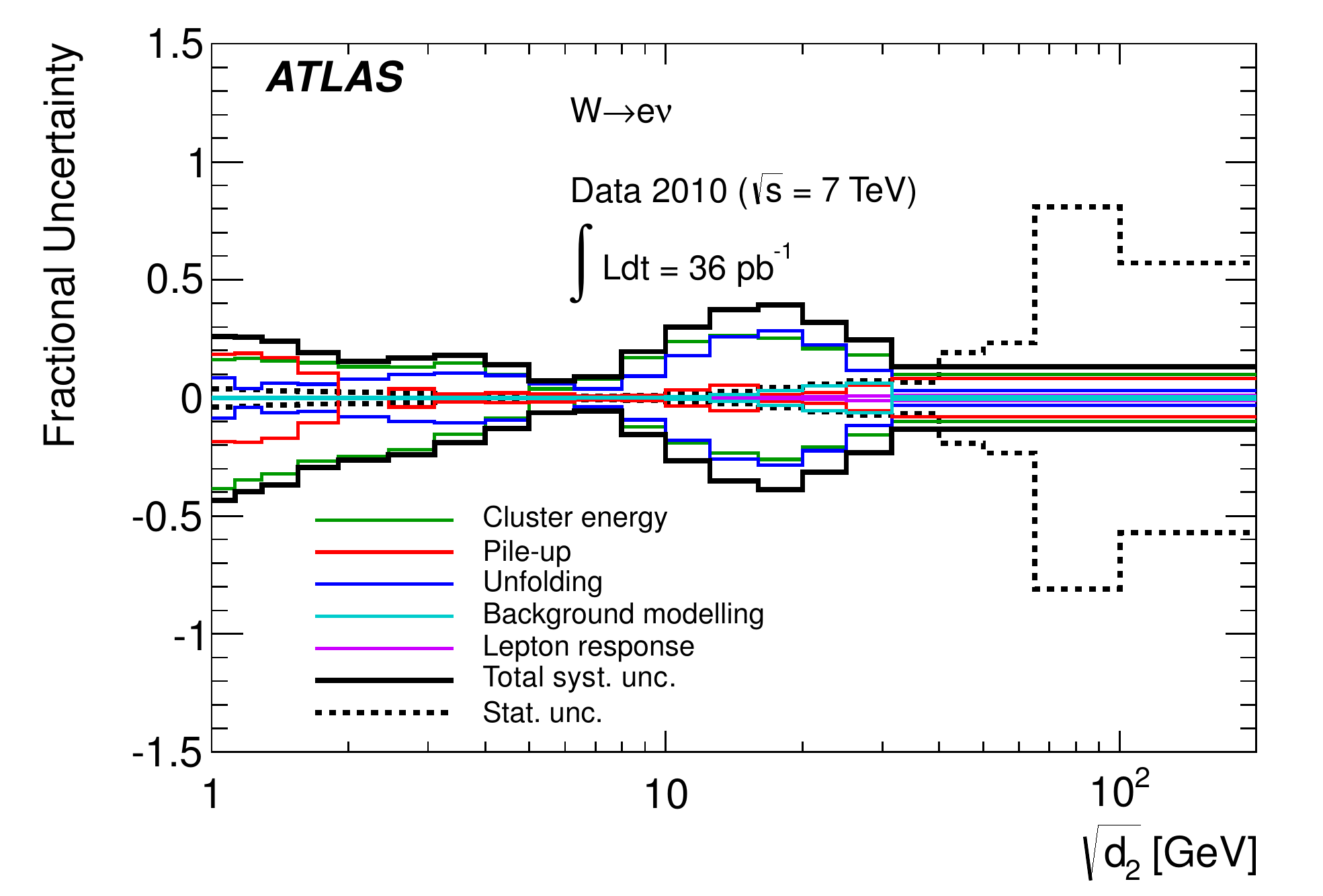}\nolinebreak
\includegraphics[width=0.5\textwidth]{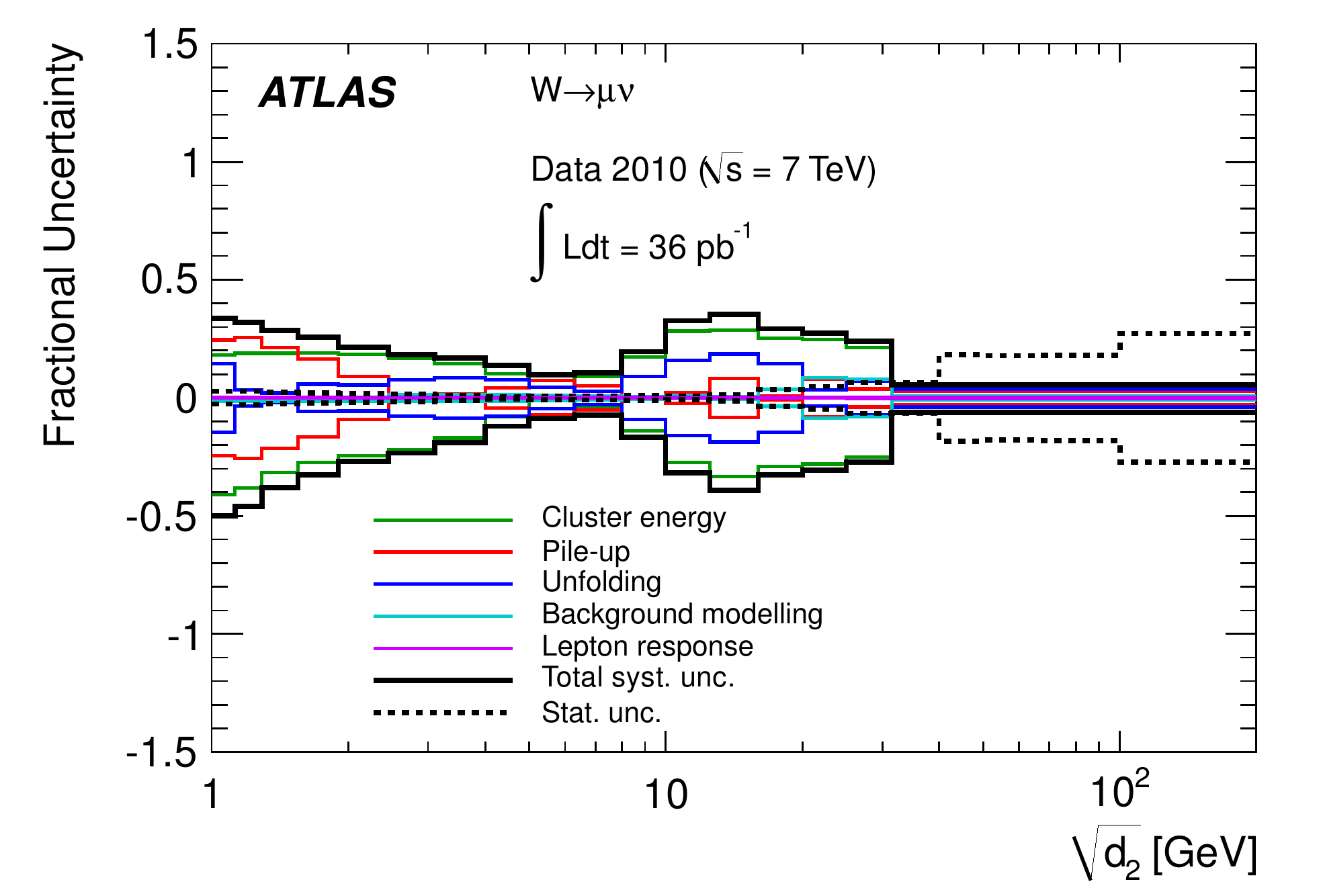}\\
\includegraphics[width=0.5\textwidth]{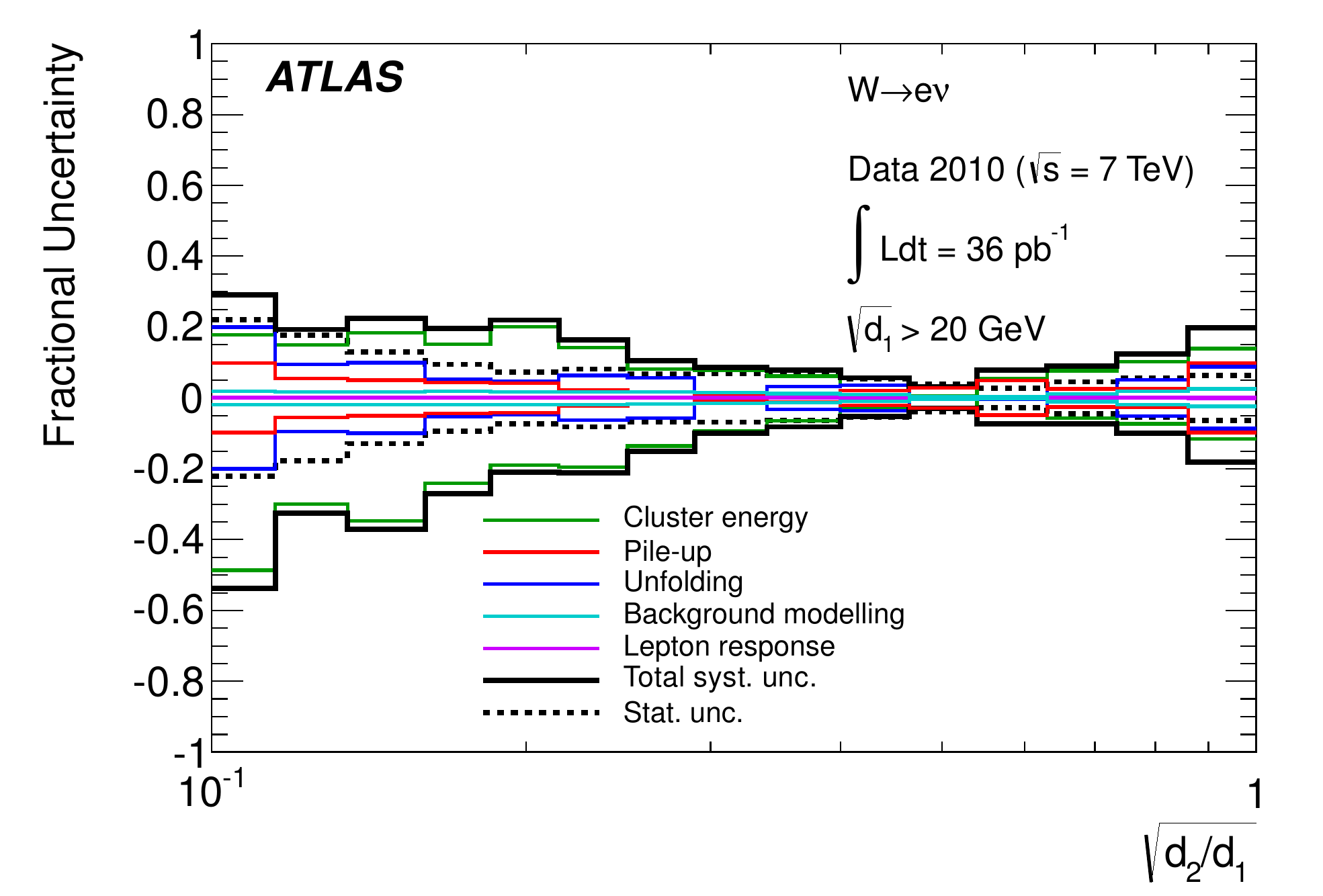}\nolinebreak
\includegraphics[width=0.5\textwidth]{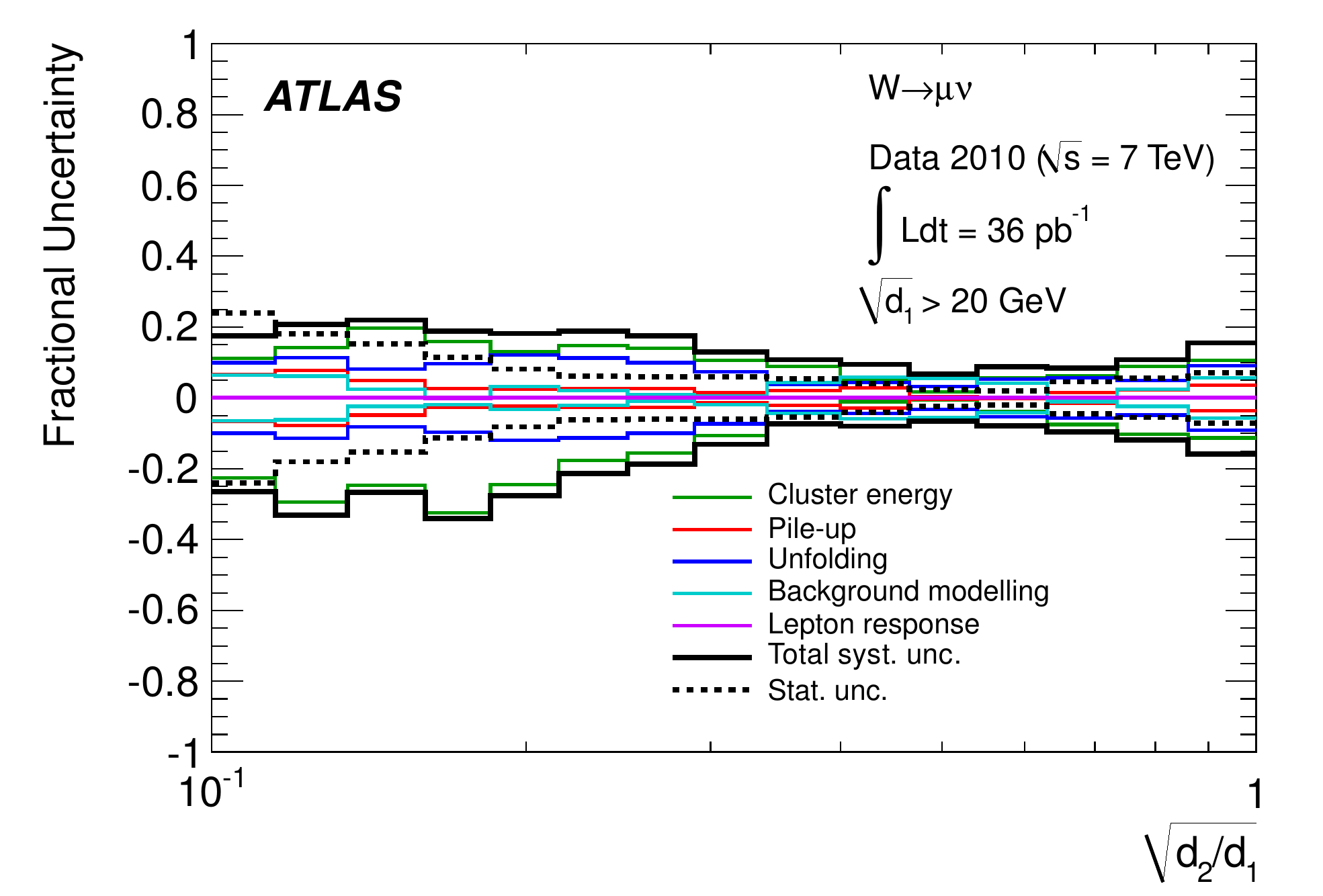}
\caption{Summary of the systematic uncertainties on the measured particle-level distributions for $\sqrt{d_{1}}$ (top) and $\sqrt{d_{2}}$ (middle) and the ratio $\sqrt{d_{2}/d_{1}}$ (bottom) in the $W\to e\nu$ (left) and $W\to \mu\nu$ (right) channels.}
\label{figure_overallunc_d12}
\end{center}
\end{figure}

\clearpage
\begin{flushleft}
{\Large The ATLAS Collaboration}

\bigskip

G.~Aad$^{\rm 48}$,
T.~Abajyan$^{\rm 21}$,
B.~Abbott$^{\rm 111}$,
J.~Abdallah$^{\rm 12}$,
S.~Abdel~Khalek$^{\rm 115}$,
A.A.~Abdelalim$^{\rm 49}$,
O.~Abdinov$^{\rm 11}$,
R.~Aben$^{\rm 105}$,
B.~Abi$^{\rm 112}$,
M.~Abolins$^{\rm 88}$
% ...
\bigskip
\\
% ...
$^{11}$ Institute of Physics, Azerbaijan Academy of Sciences, Baku, Azerbaijan\\
$^{12}$ Institut de F{\'\i}sica d'Altes Energies and Departament de F{\'\i}sica de la Universitat Aut{\`o}noma de Barcelona, Barcelona, Spain\\
% ...
$^{21}$ Physikalisches Institut, University of Bonn, Bonn, Germany\\
% ...
$^{48}$ Fakult{\"a}t f{\"u}r Mathematik und Physik, Albert-Ludwigs-Universit{\"a}t, Freiburg, Germany\\
$^{49}$ Section de Physique, Universit{\'e} de Gen{\`e}ve, Geneva, Switzerland\\
% ...
$^{88}$ Department of Physics and Astronomy, Michigan State University, East Lansing MI, United States of America\\
% ...
$^{105}$ Nikhef National Institute for Subatomic Physics and University of Amsterdam, Amsterdam, Netherlands\\
% ...
$^{111}$ Homer L. Dodge Department of Physics and Astronomy, University of Oklahoma, Norman OK, United States of America\\
$^{112}$ Department of Physics, Oklahoma State University, Stillwater OK, United States of America\\
% ...
$^{115}$ LAL, Universit{\'e} Paris-Sud and CNRS/IN2P3, Orsay, France\\
% ...
\end{flushleft}


\end{document}